\documentclass[11pt,a4paper]{article}
\usepackage{arxiv}
\usepackage[sort,comma,authoryear,round]{natbib}

\usepackage{graphicx}
\usepackage{amsthm,amsmath,amsfonts,amssymb}
\usepackage{enumitem}
\usepackage{verbatim}
\usepackage{rotating}
\usepackage{setspace}
\usepackage[nottoc]{tocbibind}
\usepackage[dvipsnames]{xcolor}
\usepackage{soul}

\definecolor{blackcolor}{rgb}{0, 0, 0}
\colorlet{purple}{blackcolor}

\usepackage{hyperref}
\hypersetup{
    colorlinks=true,     
    allcolors=blue,
}

\usepackage{bm, float, comment}
\usepackage{multirow}
\usepackage{substr}
\usepackage[dvipsnames]{xcolor}
\usepackage{makecell}

\newcommand{\checkname}[1]{%
    \IfSubStringInString{\detokenize{ARXIV}}{\jobname}{}{#1}%
}

\newcommand{\myParagraph}[1]{\paragraph*{\textit{#1}}}

\theoremstyle{definition}
\newtheorem{example}{Example}[section]

\graphicspath{{./plots/}}

\newcommand{\NNN}{{\sf I\kern-0.14emN}}   
\newcommand{\ZZZ}{{\sf Z\kern-0.45emZ}}   
\newcommand{\QQQ}{{\sf C\kern-0.48emQ}}   
\newcommand{\RRR}{{\sf I\kern-0.14emR}}   

\newcommand{\II}{{\bf I}}

\newcommand{\XX}{{\bf X}}

\newcommand{\ZZ}{{\bf Z}}

\newcommand{\boldmu}{\bm{\mu}}

\newcommand{\boldgamma}{\bm{\gamma}}

\newcommand{\boldSigma}{\bm{\varSigma}}

\newcommand{\boldeta}{\bm{\eta}}

\newcommand{\bolddelta}{\bm{\delta}}

\newcommand{\boldvtheta}{\bm{\vartheta}}
\newcommand{\boldOmega}{\bm{\Omega}}
\newcommand{\boldGamma}{\bm{\Gamma}}

\newcommand{\boldTheta}{\bm{\Theta}}
\newcommand{\boldalpha}{\bm{\alpha}}

\newcommand{\ones}{{\bf 1}}
\newcommand{\zeros}{{\bf 0}}
\newcommand{\Cor}{{\rm \mathbb{C}or}}

\newcommand{\Var}{{\rm \mathbb{V}ar}}
\newcommand{\E}{{\rm \mathbb{E}}}

\newcommand{\uu}{{\bf u}}
\renewcommand{\ss}{{\bf s}}
\newcommand{\DD}{{\bf D}}

\newcommand{\WW}{{\bf W}}

\newcommand{\MM}{{\bf M}}

\newcommand{\Norm}{\ensuremath{\mathcal{N}}}
\newcommand{\IG}{\ensuremath{\mathcal{IG}}}
\newcommand{\IW}{\ensuremath{\mathcal{IW}}}

\newcommand{\Poi}{\ensuremath{\mathcal{P}oi}}

\newcommand{\ns}{{N}}
\newcommand{\nt}{{T}}

\newcommand{\pmeas}{{p_m}}
\newcommand{\iid}{\stackrel{iid}{\sim}}
\newcommand{\half}{{1/2}}
\newcommand{\halft}{{1/2 \prime}}
\newcommand{\boldtilde}[1]{\ensuremath{\mathbf{\Tilde{#1}}}}

\newcommand{\pmtwopfive}{{PM\textsubscript{2.5}}}

\title{On the \pmtwopfive\ -- Mortality Association: A Bayesian Model for Spatio-Temporal Confounding}

\author{
 	Carlo Zaccardi \\
 	Department of Economics\\
	University G. d'Annunzio of Chieti-Pescara\\
	Pescara, Italy \\
  \texttt{carlo.zaccardi@unich.it} \\
  \And
	Pasquale Valentini \\
 	Department of Economics\\
	University G. d'Annunzio of Chieti-Pescara\\
	Pescara, Italy \\
  \And
	Luigi Ippoliti \\
 	Department of Economics\\
	University G. d'Annunzio of Chieti-Pescara\\
	Pescara, Italy \\
	\And
	Alexandra M. Schmidt \\
  Department of Epidemiology, Biostatistics and Occupational Health\\
  McGill University\\
  Montréal, Canada \\
}

\begin{document}

\maketitle


\begin{abstract}
	In epidemiological studies of air pollution and public health, estimating the health impact of exposure to air pollution may be hindered by the unknown functional form of the exposure--outcome association and by unmeasured confounding factors that are linked to both exposure and outcome. These challenges are especially relevant in spatio-temporal analyses, where their joint exploration remains limited. To study the effects of fine particulate matter on mortality among elderly people in Italy, we propose a Bayesian spatial dynamic generalized linear model that captures the non-linear exposure--outcome association and decomposes the exposure effect across fine and coarse spatio-temporal scales of variation. Together, these features allow reducing the spatio-temporal confounding bias and recovering the shape of the association, as demonstrated through simulation studies. The real-data analysis reveals a clear temporal pattern in the exposure effect, with peaks during summer months. We argue that this finding may be due to interactions of particulate matter with air temperature and unmeasured confounders.
\end{abstract}

%




%

\section{Introduction} \label{sec:tvc-introduction}

Air pollution is a major environmental exposure that increases the risk for cardiovascular and respiratory disease, among others. There is a very rich literature demonstrating the short- and long-term consequences of air pollution on health outcomes (e.g.,\citet{dominici2002air}, \citet{janes2007trends}, \citet{zeger2008mortality}, \citet{greven2011approach}, \citet{dai2014associations}, \citet{di2017association}, \citet{wu2020evaluating}, \citet{brunekreef2021mortality}, \citet{renzi2021acute}, \citet{dominici2022assessing}, \citet{gamerman2022dynamic}, \citet{requia2024short}). Collectively, these health effects translate into a sizeable contribution to population-level morbidity and premature mortality. Though the impacts of air pollution on public health are well established, quantifying those impacts on a specific population is challenging. Exposure and risk are inherently spatio-temporal phenomena, where variations in particulate types and levels, the heterogeneity in population characteristics and dynamics, as well as the context in which measurements occur all influence and complicate both estimation and interpretation.


The ultimate goal of this paper is to estimate the small-scale spatio-temporal association between all-cause mortality and air pollution for two Italian regions, Piemonte and Lombardia, over the period from 2018 to 2022 (see Section \ref{sec:tvc-motivating-example}). We limit our study to particulate matter with a diameter less than $2.5 \mu m$ (PM\textsubscript{2.5}), which is known to pose the greatest health risk.

As with any regression modeling approach, we must be concerned about the risk of potential bias arising from model misspecification. In the context of our spatio-temporal regression, the exposure--outcome relationship is challenging to model because both the exposure and the health risk vary across time and space. For example: (i) the physical properties and chemical composition of the PM mixture exhibit substantial variation across locations and over time, influenced by environmental conditions, human activities, and episodic phenomena such as sandstorms and wildfires \citep{chen2013seasonal,kelly2012size, masselot2022differential, peng2005seasonal, zhou2021excess}; (ii) human susceptibility to \pmtwopfive\ is not static, but is modulated by temporally and spatially varying factors, including extreme temperatures, which can exacerbate health risks in specific regions or seasons \citep{bergmann2020effect,peng2005seasonal,roberts2004interactions,li2017modification}; (iii) the relationship between true and measured exposure varies across time and space, with seasonal factors --- such as increased summer ventilation --- potentially reducing exposure measurement error \citep{bergmann2020effect,peng2006model,roberts2004interactions,stafoggia2008does}; (iv) factors such as lockdowns, the use of facial masks, physical distancing and other COVID-19 related measures have significantly altered people's habits and interactions with the environment since 2020; and (v) morphological and climatic differences across health districts may influence the exposure--outcome relationship.

Often the simplifying assumption of a constant and linear exposure--outcome relationship is made.  However, if interactions with confounding variables are ignored, this simple relationship may be inadequate. We acknowledge that all models are inherently flawed, but some of them are so misspecified that they can lead to incorrect inferential statements \citep{gelman2013philosophy}.

There are several established strategies for incorporating a non-linear exposure--outcome relationship. One option is to adopt a functional approach and estimate an exposure--response function \citep[e.g.,][]{brunekreef2021mortality,cork2023methods,dominici2022assessing,renzi2021acute}. However, the inherent spatial and temporal variation in the relationship between \pmtwopfive\ and mortality is typically not accounted for. Alternatively, and the strategy used in this work, one can specify a model that explicitly incorporates spatial and temporal variation in the relationship of interest \citep{gelfand2003spatial}. 

Within the second strategy, several modeling approaches have been proposed; however, they present limitations within our framework. For example, the exposure effect can be modeled either as a periodic function \citep{chen2013seasonal,peng2005seasonal} or as a season-specific step function \citep{dai2014associations, klompmaker2023effects}. However, because the shape of the effect is assumed to be the same across years, these models cannot account for exceptional events such as the outbreak of the SARS-CoV-2 pandemic, which occurred during the period of interest in our study. Further, the abrupt changes in the exposure effect between adjacent seasons implicit in the approaches proposed by \citet{dai2014associations} and \citet{klompmaker2023effects} are unlikely in practice. The exposure effect depends on meteorological conditions and human activity patterns, which tend to change smoothly over time \citep{peng2005seasonal,bhaskaran2013time}. \citet{chiogna2002dynamic} and \citet{lee2008modelling} employ dynamic generalized linear models \citep[DGLMs;][]{west1997bayesian} to analyze the relationship between air pollution and mortality in time series studies; however, their methods are not suitable for spatio-temporal data. Spatio-temporally varying coefficients models \citep[STVCMs;][]{banerjee2014hierarchical, sahu2022bayesian} have also been applied in a variety of fields, including healthcare \citep{cai2013bayesian, song2020spatiotemporally}, as well as air pollution and climate studies \citep{paez2008spatially, bakar2015spatiodynamic}. However, these models do not incorporate space-time interaction terms, thus limiting their ability to capture complex spatio-temporal patterns.

In addition to model misspecification bias, we also face potential bias from confounding. When data are spatially correlated, the challenge has received considerable attention in recent years --- e.g., \citet{reich2006effects}, \citet{paciorek2010importance}, \citet{thaden2018structural}, \citet{papadogeorgou2019adjusting}, \citet{keller2020selecting}, \citet{giffin2021instrumental}, \citet{gilbert2021causal}, \citet{nobre2021effects}, \citet{bobb2022accounting}, \citet{dupont2022spatial+}, \citet{khan2022restricted}, \citet{marques2022mitigating}, \citet{zimmerman2022deconfounding}, \citet{azevedo2023alleviating}, \citet{guan2023spectral}, \citet{woodward2024instrumental}, \citet{gilbert2025consistency}, \citet{wiecha2025two}, \citet{wu2025spatial} and \citet{zaccardi2025regularized}. \citet{gilbert2021causal} and \citet{khan2023re} note that the challenge of ``spatial confounding'' is further complicated by the lack of a single, common definition, and summarized several phenomena related to spatial confounding found in the literature.

Though the phenomena summarized in \citet{gilbert2021causal} and \citet{khan2023re} can be naturally extended to data that are structured in both space and time, the problem of \textit{spatio-temporal confounding} has received comparatively less attention. \citet{reich2021review} discuss extensions of causal inference methods to settings where exposures and outcomes vary across space and time. \citet{prates2022non} and \citet{adin2023alleviating} propose extensions of methods originally developed for purely spatial settings, inspired by the restricted regression approach \citep{reich2006effects, hodges2010adding}. Their proposed methods adopt the ``random effect collinearity'' perspective, where confounding arises from collinearity between exposure and structured random effects \citep{gilbert2021causal}. 

In this work, we adopt an alternative definition, that is, spatio-temporal confounding is taken as ``omitted confounder bias,'' which refers to the existence of unmeasured confounding factors that have a spatio-temporal structure affecting both exposure and outcome \citep{gilbert2021causal}. In epidemiological studies on the health effects of air pollution, information on some confounders --- such as meteorological variables --- is typically available. However, others may be unmeasured or difficult to quantify, such as influenza epidemics or changes in population structure \citep{peng2006model}. Failing to fully account for unmeasured confounding can lead to an incorrect inference about the exposure--outcome relationship (due to confounding bias).

We seek a comprehensive data-driven approach that is able to 1) adequately control for spatio-temporal confounding; 2) capture the pattern of the exposure effect in space and/or time; and 3) break down the spacetime-varying coefficients into simpler parts to enhance interpretability. We, therefore, propose the \textit{spatial DGLM for confounding (SDGLMC)}. To implement this structure, we adopt a Bayesian DGLM framework introducing spacetime-varying intercept and slope coefficients into the model specification to account for unmeasured spatio-temporal confounders and non-linear exposure--outcome associations, respectively. Because DGLMs are only \textit{locally} linear, they can approximate any smooth function of time \citep{gamerman2006markov,granger2008non}. Further, because DGLMs capture gradual changes and emphasize both seasonal and non-systematic variations \citep{lutkepohl2005new, petris2009dynamic, bitto2019achieving}, SDGLMC can accommodate exceptional events and avoids abrupt shifts in the exposure effect. Following \citet{greven2011approach} and \citet{janes2007trends}, we decompose the exposure effect into two orthogonal components, each of which describes the (non-linear) association with the health outcome at a different scale of variation (see Section \ref{sec:tvc-model}). The focus of inference is the component that captures the association at a finer scale --- where confounding by unmeasured factors is less likely \citep{greven2011approach} --- and is assumed to exhibit spatio-temporal dynamics. Furthermore, SDGLMC enables explicit interpretation of each component of these dynamics (i.e., baseline, temporal, spatial, and interaction effects) through an analysis-of-variance-like decomposition (see Section \ref{subsec:tvc-coefficients}).

This paper is organized as follows. Section \ref{sec:tvc-motivating-example} offers a brief description of the data that motivated this study. Sections \ref{sec:tvc-model} and \ref{sec:tvc-inference} provide detailed information about the proposed model and the inferential procedure, respectively. Section \ref{sec:tvc-competing-approaches} reviews existing approaches from the literature, which serve as benchmarks for comparison with the results of the proposed model. In Section \ref{sec:tvc-simulations}, simulation studies are conducted to evaluate the ability of SDGLMC to control for spatio-temporal confounding and recover the true association. The results of the real-data application are presented in Section \ref{sec:tvc-application}. Finally, Section \ref{sec:tvc-conclusions} provides insights from the analysis and offers a concluding discussion.

\section{Motivation} \label{sec:tvc-motivating-example}

We analyze the association between daily \pmtwopfive\ concentrations and all-cause mortality counts among people aged 65 and older in the Italian regions of Piemonte and Lombardia between 2018 and 2022.


The data were retrieved from publicly available sources; the data sources and processing procedures are detailed in Section A of the Supplementary Material. Mortality counts were collected from the Italian National Institute of Statistics \citep{istat_decessi_comuni} and are observed on an irregular lattice comprising $\ns=117$ non-overlapping areal units, termed \textit{health districts} (HDs). We denote the number of deaths (by any cause) that occurred in the $i$-th HD, $i=1,\dots,\ns$, on day $t=1,\dots,\nt$ as $Y_{it}$, where $\nt$ is the study period  ($\nt=1826$ consecutive days). \pmtwopfive\ concentrations are used as the measure of exposure in HD$_i$ on day $t$, denoted as $X_{it}$, and were retrieved from the Copernicus Atmosphere Monitoring Service's (CAMS) Atmosphere Data Store \citep{CAMSdataset,copernicus2023cams}. Temperature and relative humidity measurements were obtained from the ERA5 reanalyses dataset available at the Copernicus Climate Change Service's (C3S) Climate Data Store \citep{copernicus2023era5,hersbach2023era5} and were spatially and temporally aggregated to match the resolution of the health data.

\begin{figure}[!t]
  \centering
  \begin{minipage}{0.48\textwidth}
      \includegraphics[width=\textwidth]{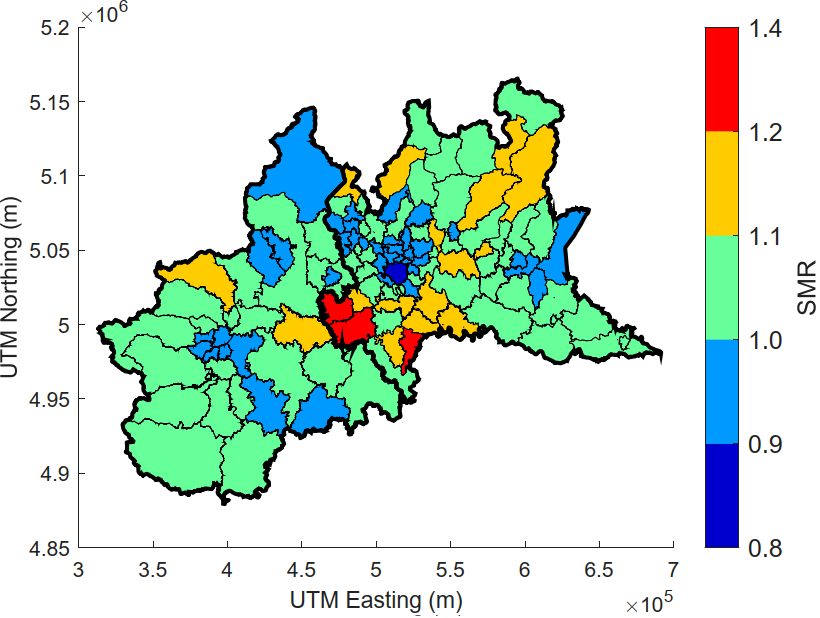}
  \end{minipage}
  \hfill
  \begin{minipage}{0.48\textwidth}
      \includegraphics[width=\textwidth]{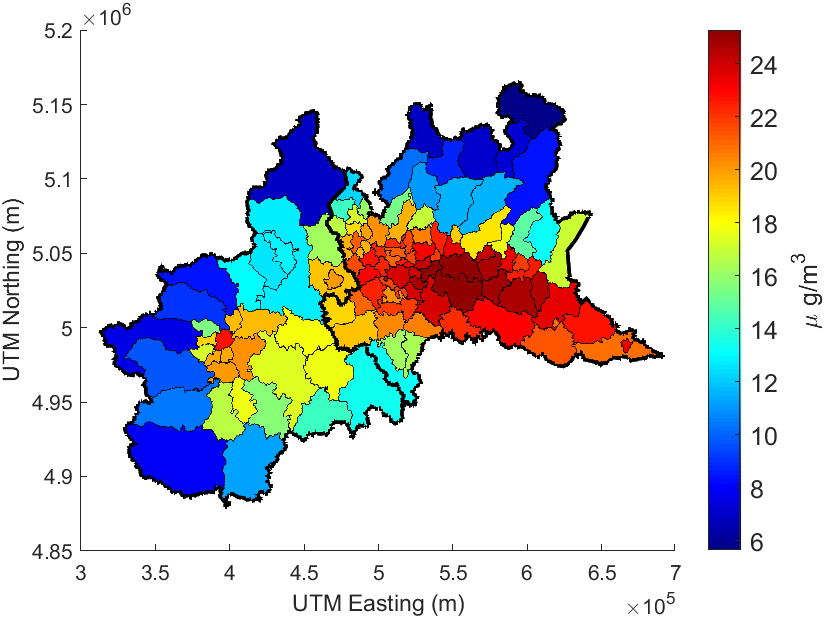}
  \end{minipage}
  \caption[Long-term averages of the SMR and the \pmtwopfive\ levels]{Long-term averages of SMR (left) and \pmtwopfive\ concentrations (right) across HDs from 2018 to 2022. The thicker solid line separates Piemonte (the westernmost region) from Lombardia (the easternmost region).}
  \label{fig:tvc-y-x-space}
\end{figure}

A measure commonly employed for exploratory analyses is the standardized mortality ratio (SMR), which is calculated as $SMR_{it} = Y_{it}/E_{it}$, for $i=1,\dots,\ns$ and $t=1,\dots,\nt$. The denominator, $E_{it}$, denotes the expected number of deaths and is computed allowing for the underlying age--sex stratification of the population within each HD (see Section A of the Supplementary Material). The daily SMR for each HD averaged across the study period is shown in the left panel of Figure \ref{fig:tvc-y-x-space}, where noticeable spatial clustering is seen. The mean SMR across the study domain is $1.25$. The right panel of Figure \ref{fig:tvc-y-x-space} shows the long-term average \pmtwopfive\ concentrations for each HD, indicating that Lombardia is generally more polluted than Piemonte ($19.41 \, \mu g/m^3$ and $15.18 \, \mu g/m^3$, respectively). Additional descriptive statistics are provided in Section B of the Supplementary Material.

\begin{figure}[!htbp]
  \centering
  \includegraphics[clip,width=14cm]{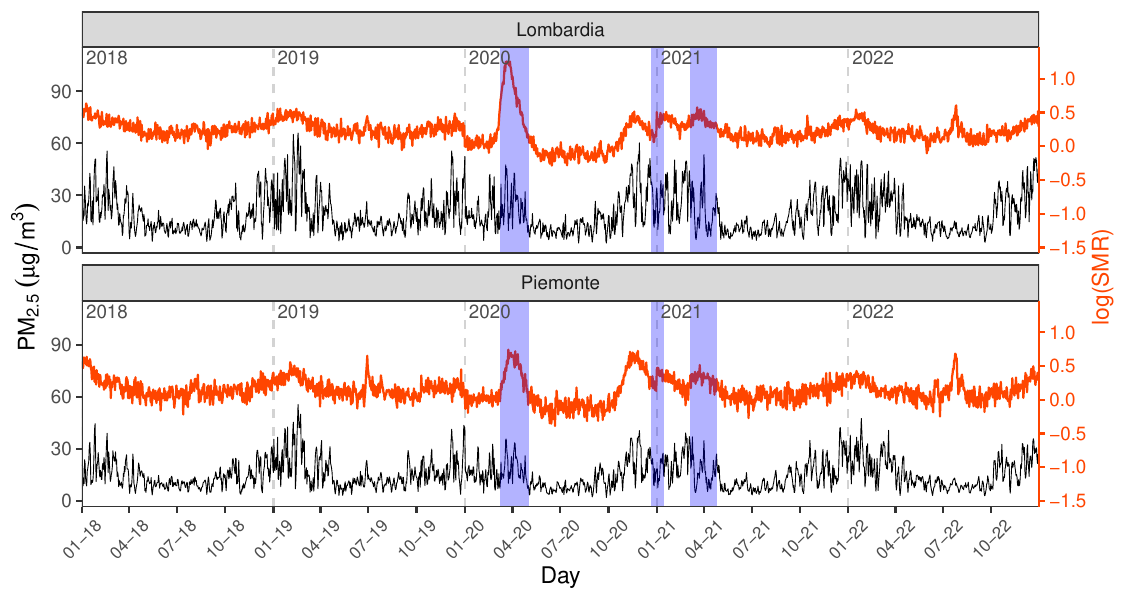}
  \caption[Time series of the average daily log(SMR) and pollution]{Time series of log-transformed daily SMR (orange) and \pmtwopfive\ concentrations (black) for Piemonte (top) and Lombardia (bottom), spanning 2018--2022. Vertical dashed lines mark calendar years. orange bands indicate national lockdown periods.}
  \label{fig:tvc-smr-regions}
\end{figure}

Figure \ref{fig:tvc-smr-regions} shows the observed daily average SMR on the log scale (orange) and the daily average \pmtwopfive\ concentration (black) for Lombardia and Piemonte. Both regions exhibit similar seasonal patterns, with the lowest average SMR and \pmtwopfive  values in summer --- except for the year 2020. Two distinct SMR peaks are observed in 2020, one in April and another in November, corresponding to the first two waves of the SARS-CoV-2 pandemic. The first wave was more severe in Lombardia, while the second one was more pronounced in Piemonte. Notably, these peaks in the SMR time series are not reflected in the \pmtwopfive\ time series. In fact, as shown in Table B.3 of the Supplementary Material, the lowest overall annual mean \pmtwopfive\ during the study period was observed in 2020.

Additional features can be seen in Figure \ref{fig:tvc-smr-regions}. For example, the highest SMR values in 2018 occurred in early January, suggesting that the seasonal influenza outbreak reached its peak before the end of 2017. In contrast, the SMR peaked again in February of the following year, suggesting a longer or later influenza outbreak. The increase in mortality observed during summer 2022 may be due to the occurrence of prolonged heat waves --- the average daily temperature time-series has its absolute maximum in July 2022 --- or an uncommon summer influenza outbreak.

Figure \ref{fig:tvc-smr-regions} illustrates the complexity of SMR dynamics over time. Some features clearly reflect factors unrelated to \pmtwopfive. The departures highlight the need for models that can account for the effects of measured and unmeasured confounders. Failing to do so may lead to incorrect inferences. The modeling approach proposed in the next section is designed to address this issue.

\section{The Proposed Model} \label{sec:tvc-model}

It is assumed that the outcome, $Y_{it}$, follows a Poisson distribution with mean $\mu_{it}$, for $i=1,\dots,\ns$ and $t=1,\dots,\nt$. The linear predictor, $\vartheta_{it} = \ln (\mu_{it})$, is modeled as a function of exposure $X_{it}$, measured confounders $\mathbf{M}_{it} = ({M}_{1it}, \dots, {M}_{\pmeas it})'$, and an unmeasured confounder $Z_{it}$, which is smooth in space and time.
Assuming that ${\mathbf{M}}_{it}$ and $Z_{it}$ are spatially and temporally smoother than the exposure, the model is specified as follows:
\begin{align}
  Y_{it}|\mu_{it} &\stackrel{ind}{\sim} \Poi (\mu_{it}) \label{eq:tvc-outcome} \\
  \vartheta_{it} &= \ln(\mu_{it}) = o_{it} + \beta_{0} + f(X_{it}; \boldgamma_f) + {\mathbf{M}}_{it}' \boldalpha + Z_{it} + u_{it}\,, \label{eq:tvc-linear-predictor}
\end{align}
where 
$o_{it} = \log(E_{it})$ is the offset, $\beta_{0}$ is an intercept, $f(\cdot; \cdot)$ is an unknown smooth function parameterized by the vector $\boldgamma_f$, $\boldalpha$ is a $\pmeas$-dimensional vector of coefficients, and $u_{it}$ is a spatially and temporally uncorrelated Gaussian error term with zero mean and constant variance, $u_{it} \stackrel{iid}{\sim} \Norm (0, \tau_{u}^2)$.

If all confounders were measured and $f(\cdot; \cdot)$ was modeled using basis functions, the model specification in Equations (\ref{eq:tvc-outcome}) and (\ref{eq:tvc-linear-predictor}) would belong to the class of generalized additive models \citep[GAMs; see][]{wood2017generalized}. Instead, we consider that $f(\cdot; \cdot)$ can be adequately approximated by the first-order Taylor's expansion:
\begin{equation} \label{eq:tvc-taylor}
    f(X_{it}; \boldgamma_f) = f(\widehat{X}_{it}; \boldgamma_f) + f'(\widehat{X}_{it}; \boldgamma_f) (X_{it} - \widehat{X}_{it}) + \varpi_{it}\,,
\end{equation}
where $\widehat{X}_{it} $ denotes the large-scale (or \textit{overall}) spatio-temporal trend, $f'(\widehat{X}_{it}; \boldgamma_f)$ is the first derivative of $f(\cdot; \boldgamma_f)$ with respect to its first argument evaluated at $\widehat{X}_{it}$, and $\varpi_{it}$ denotes the Taylor's remainder. To compute the spatio-temporal trend, $X_{it}$ can be regressed on a smooth function of space and time \citep[see also][]{janes2007trends}:
\begin{equation} \label{eq:tvc-Xmean}
  X_{it} = h(\ss_i, t; df) + \nu_{it}\,,
\end{equation}
where $\ss_i$ denotes the spatial coordinates of the centroid of areal unit $i$, $df$ are the degrees of freedom of the smooth term, and $\nu_{it}$ is an uncorrelated Gaussian random variable with zero mean and constant variance. The overall trend is then defined as the fitted values obtained from the model (Equation (\ref{eq:tvc-Xmean})), i.e., $\widehat{X}_{it} = \widehat{h}(\ss_i, t; df)$.

Equation (\ref{eq:tvc-taylor}) thus decomposes the exposure effect into two orthogonal components: one at large scale of variation - $f(\widehat{X}_{it}; \boldgamma_f)$ - and one at a small (or \textit{local}) scale - $f'(\widehat{X}_{it}; \boldgamma_f) (X_{it} - \widehat{X}_{it})$. This decomposition helps disentangle the effect of interest from that of unmeasured confounders. As discussed in the econometric and spatio-temporal literatures, the use of detrended exposure helps remove large-scale spatio-temporal variation, thereby reducing the risk of spurious associations driven by underlying trends \citep{Wooldridge2020}. At the same time, it preserves small-scale spatio-temporal fluctuations. From an epidemiological point of view, $\widehat{X}_{it}$ captures trend and seasonal components of the exposure and, thus, is expected to be highly correlated with $Z_{it}$ (which is assumed to be smooth), whereas the correlation between $X_{it} - \widehat{X}_{it}$ and $Z_{it}$ should be close to zero.
Therefore, the unmeasured confounder conceivably affects the association between outcome and exposure at large scales \citep{greven2011approach,janes2007trends}, so the first term in Equation (\ref{eq:tvc-taylor}) can potentially be biased, but it is less likely that the second term $f'(\widehat{X}_{it}; \boldgamma_f) (X_{it} - \widehat{X}_{it})$ is affected by confounding bias.

Substituting the decomposition into Equation (\ref{eq:tvc-linear-predictor}), we obtain:
\begin{equation*}
    \vartheta_{it} = o_{it} + \beta_{0} + f(\widehat{X}_{it}; \boldgamma_f) + f'(\widehat{X}_{it}; \boldgamma_f) (X_{it} - \widehat{X}_{it}) + {\mathbf{M}}_{it}' \boldalpha + Z_{it} + \tilde{u}_{it} \,, 
\end{equation*}
where $\tilde{u}_{it} = u_{it} + \varpi_{it}$ is such that $\tilde{u}_{it} \iid \Norm(0, \tilde{\tau}_u^2)$.

For each $i$, the unknown function $f(\widehat{X}_{it}; \boldgamma_f)$ and its first derivative can be seen as functions of space and time, and spacetime-varying coefficients can adequately represent them \citep[see also][Section 3.1]{west1997bayesian}. Therefore, we propose the following reparameterization:
\begin{equation} \label{eq:tvc-latentBETA}
    \vartheta_{it} = o_{it} + \beta_{0it} + \beta_{1it} (X_{it} - \widehat{X}_{it}) + {\mathbf{M}}_{it}' \boldalpha + \tilde{u}_{it}\,,
\end{equation}
where $\beta_{1it}=f'(\widehat{X}_{it}; \boldgamma_f)$ is the spacetime-varying coefficient representing the local association between exposure and outcome. It denotes the association between day-to-day variation in area-specific deviations in \pmtwopfive\ concentrations from the large-scale spatio-temporal trend, and the corresponding day-to-day variation in area-specific deviations in mortality rates from the large-scale trend. The parameter $\beta_{0it}$ is a spacetime-varying intercept that accounts for both the large-scale exposure effect and the unmeasured confounder:
\begin{equation*}
  \beta_{0it} = \beta_{0} + f(\widehat{X}_{it}; \boldgamma_f) + Z_{it}\,.
\end{equation*}

The following examples illustrate that under certain function forms of $f(\cdot; \cdot)$, our model extends other methods proposed in the literature.

\begin{example} \label{ex:tcv-1}
  If $f(\cdot; \boldgamma_f)$ is a constant function, i.e., $f(X_{it}; \boldgamma_f) = \gamma_{f0}$, then $\beta_{1it} = 0$ and $\beta_{0it} = \beta_{0} + \gamma_{f0} + Z_{it}$.
\end{example}

\begin{example} \label{ex:tcv-2}
  If $f(\cdot; \boldgamma_f)$ is a linear function, i.e., $f(X_{it}; \boldgamma_f) = \gamma_{f0} + \gamma_{f1} X_{it}$, then $\beta_{1it} = \gamma_{f1}$ and $\beta_{0it} = \beta_{0} + \gamma_{f0} + \gamma_{f1}\widehat{X}_{it} + Z_{it}$. The model proposed by \citet{greven2011approach} can be cast in this case, with $\gamma_{f0}=0$. The authors, however, assume further that the intercepts do not depend on time. 
\end{example}


\subsection{Specification of the Spacetime-Varying Coefficients} \label{subsec:tvc-coefficients}

Although it is possible to model $\beta_{0it}$ and $\beta_{1it}$ directly, we decompose them to enhance prior elicitation and interpretability \citep[see also][]{bakar2015spatiodynamic,cai2013bayesian,paez2008spatially}:
\begin{equation*} \label{eq:tvc-coefficients}
    \beta_{kit} = \overline{\delta}_k + \tilde{\delta}_{ki} + \delta_{kt} + \delta_{kit}^* \,, \quad k=0,1\,,
\end{equation*}
where $\overline{\delta}_k$ is the baseline effect of the $k$th covariate, which is constant across space and time; $\tilde{\delta}_{ki}$ is the area-specific effect of the $k$th covariate; $\delta_{kt}$ is the time-specific effect of the $k$th covariate; and $\delta_{kit}^*$ is the spatio-temporal interaction effect of the $k$th covariate. For each $i$, $\tilde{\delta}_{ki}$ represents the average (across time) spatial deviation from the baseline effect. In particular, $\tilde{\delta}_{1i}$ captures time-independent spatial heterogeneity in the exposure--outcome association at the small scale of variation (recall that the regressor is $X_{it} - \widehat{X}_{it}$). Thus, the parameter $\tilde{\delta}_{1i}$ can inform about \textit{where} populations are most vulnerable to air pollution.

For each time $t$, $\delta_{kt}$ can be interpreted as the average (across space) temporal deviations from the baseline effect. Thus, $\delta_{1t}$ represents the temporal evolution of the association of interest, and highlights \textit{when} peaks can occur.

The interaction effect, $\delta_{kit}^*$, captures the area- and time-specific deviations that are left after accounting for the temporal and spatial effects. The simulation scenarios in Section \ref{sec:tvc-simulations} are built based on plausible values of $\beta_{1it}$, which are typically very small in empirical applications. Thus, given the tiny contribution of the term $\beta_{1it} (X_{it} - \widehat{X}_{it})$ to the augmented-data likelihood (see Section \ref{subsec:tvc-posterior}), it is difficult to recover the interaction effect of the exposure, even in the absence of confounding. Therefore, we posit that $\delta_{1it}^*$ equals zero, an assumption consistent with prior research \citep[e.g.,][]{bakar2015spatiodynamic,paez2008spatially}.

In SDGLMC, we implement a DGLM with random walk (RW) processes to model the temporal effects (namely, $\delta_{0t}$ and $\delta_{1t}$) and the interaction effect, $\delta_{0it}^*$. We propose the following state-space formulation for the latent process and the regression coefficients, for $t=1, \dots, \nt$:
\begin{align}
  \boldvtheta_{t} &= \mathbf{o}_t + \ones_\ns (\overline{\delta}_0 + \delta_{0t}) + \Tilde{\bolddelta}_0 + \bolddelta_{0t}^* + %
  \boldtilde{X}_{t} (\overline{\delta}_1 + \delta_{1t}) + \boldtilde{X}_{t} \odot \Tilde{\bolddelta}_1 +%
  {\mathbf{M}}_t \boldalpha + \tilde{\uu}_{t} \label{eq:tvc-obs_proposal} \\
  \delta_{0t} &= \delta_{0,t-1} + w_{0t}, \quad w_{0t} \sim \Norm(0, \sigma_{w0}^2) \label{eq:tvc-state_proposal2} \\
  \bolddelta_{0t}^* &= \bolddelta_{0,t-1}^* + \mathbf{w}_{0t}^*, \quad \mathbf{w}_{0t}^* \sim ICAR(\sigma_{w0}^{* 2}) \label{eq:tvc-state_proposal1} \\
  \delta_{1t} &= \delta_{1,t-1} + w_{1t}, \quad w_{1t} \sim \Norm(0, \sigma_{w1}^2) \label{eq:tvc-state_proposal}
\end{align}

\noindent where the symbol $\odot$ denotes the Hadamard product, $\boldvtheta_t = (\vartheta_{1t}, \dots, \vartheta_{\ns t})'$, $\mathbf{o}_t = (o_{1t}, \dots, o_{\ns t})'$, $\Tilde{\bolddelta}_k = (\tilde{\delta}_{k1}, \dots, \tilde{\delta}_{k \ns})'$ are the spatial-specific effects for $k=0,1$, $\tilde{\uu}_{t} = (\tilde{u}_{1t}, \dots, \tilde{u}_{\ns t})' \sim \Norm(\zeros, \tilde{\tau}_u^2 \II_\ns)$, $\bolddelta_{0t}^* = (\delta_{01t}^*, \dots, \delta_{0 \ns t}^*)'$, and the observation matrices are as follows:
\begin{equation*}
  \boldtilde{X}_{t} = \begin{bmatrix} X_{1t} - \widehat{X}_{1t} \\ \vdots \\ X_{\ns t} - \widehat{X}_{\ns t} \end{bmatrix} \quad \text{and} \quad
  \mathbf{M}_t = \begin{bmatrix} \mathbf{M}_{1t}' \\ \vdots \\ \mathbf{M}_{\ns t}' \end{bmatrix}  \,.
\end{equation*}

Note that Equation \eqref{eq:tvc-state_proposal1} is one of five alternatives that will be discussed further in Section \ref{sec:tvc-application}. To capture the underlying spatial structure, $\Tilde{\bolddelta}_0$ and $\Tilde{\bolddelta}_1$ are modeled as intrinsic conditional autoregressive \citep[ICAR; e.g.,][]{banerjee2014hierarchical} processes, denoted as $ICAR(\sigma_{\delta k}^2) $, with probability density function proportional to
\begin{equation*} \label{eq:tvc-CAR}
  p(\Tilde{\bolddelta}_k) \propto \exp \biggl\{ - \frac{1}{2\sigma_{\delta k}^2} \Tilde{\bolddelta}_k^\prime (\mathbf{D} -  \mathbf{W}) \Tilde{\bolddelta}_k \biggr\}, \quad k=0,1 \,,
\end{equation*}
where $\mathbf{W}$ is the spatial adjacency matrix with elements $(\mathbf{W})_{ij}=1$ if areal units $i$ and $j$ are neighbors, and 0 otherwise; $\mathbf{D}$ is a diagonal matrix constructed from the sum of the rows of $\mathbf{W}$; 
and $\sigma_{\delta k}^2$ is a variance parameter. Such structure has been previously considered by \citet{franco2019unified, franco2022variance} to model spatially varying coefficients.


\section{Bayesian Inference and Computations in SDGLMC} \label{sec:tvc-inference}

The first stages of the Bayesian hierarchical model are defined by Equations (\ref{eq:tvc-outcome}) and (\ref{eq:tvc-obs_proposal})--(\ref{eq:tvc-state_proposal}). To complete these stages, we assign a prior distribution on the parameter set, $\boldTheta = \{ \overline{\delta}_0, \delta_{00}, \bolddelta_{00}^*, \overline{\delta}_1, \delta_{10}, \boldalpha, \sigma_{w0}^2, \sigma_{\delta 0}^2, \sigma_{w0}^{* 2}, \sigma_{w1}^2, \sigma_{\delta 1}^2, \tilde{\tau}_{u}^2 \}$. 

\subsection{Prior Specification} \label{subsec:prior}
Independent prior distributions are imposed on each component of $\boldTheta$. It is assumed that the initial states of the time-varying coefficients follow a Gaussian distribution, namely
%
$\delta_{k0} \sim \Norm (0, V_{\delta k})$, $k=0,1$, where $V_{\delta 0} = 4$ and $V_{\delta 1} = 4 \times 10^{-6}$. Also, $\bolddelta_{00}^* \sim \Norm (\mathbf{0}, \mathbf{V}_{\delta}^*)$, with $\mathbf{V}_{\delta}^* = 10 \mathbf{I}_\ns$.

Inverse-gamma priors are imposed on the variances of the innovation errors, namely $\sigma_{w0}^{* 2} \sim \IG (a_{w0}^*, b_{w0}^*)$, $\sigma_{wk}^2 \sim \IG (a_{wk}, b_{wk}) $ and $\sigma_{\delta k}^2 \sim \IG (a_{\delta k}, b_{\delta k})$, for $k=0,1$. For $k=0$, all the hyperparameters are assumed to be $0.01$. For $k=1$, care must be taken when specifying the hyperparameters because capturing gradual changes is a desired feature of the proposed model. If non-informative prior distributions are imposed on the variances, $\sigma_{w1}^2$ and $\sigma_{\delta 1}^2$, the variations in the corresponding temporal and spatial effects become unlikely pronounced. Therefore, a prior sensitivity analysis was conducted, where the hyperparameters $a_{w1}$, $b_{w1}$, $a_{\delta 1}$ and $ b_{\delta 1}$ were varied to obtain a posterior distribution concentrated around relatively small values. The interpretation of ``small'' in this context is contingent upon the sample size and the extent to which the prior distribution influences the posterior inference relative to the likelihood. In the case of both simulated and real data (see Sections \ref{sec:tvc-simulations} and \ref{sec:tvc-application}, respectively), we find that assuming $a_{w1}=10$, $b_{w1}=0.001$, $a_{\delta 1}=1$, and $b_{\delta 1}=0.01$ prevents overfitting as a priori $\Pr (\sigma_{w1}^2 \le 2.4 \times 10^{-4}) = 0.99$ and $\Pr (\sigma_{\delta 1}^2 \le 0.995) = 0.99$.

The baseline effects $\overline{\delta}_0$ and $\overline{\delta}_1$ and the coefficients of the measured confounders $\boldalpha$ are assumed to be independently and normally distributed with mean zero and variance $10^6$. 
Finally, we assume that $\tilde{\tau}_{u}^2 \sim \IG(1, 2.1 \times 10^{-5}) $. These hyperparameters are chosen to reflect the belief that the residual relative risks $\exp(\tilde{u}_{it})$ follow a log-\textit{t} distribution with two degrees of freedom, and with $95\%$ of prior mass within the interval $(0.98, 1.02)$, as suggested by \citet{wakefield2007disease}.

\subsection{Computations} \label{subsec:tvc-posterior}
The inference procedure starts by first estimating the overall temporal trend $\widehat{X}_{it}$ as the fitted values from Equation (\ref{eq:tvc-Xmean}), where $h(\ss_i, t; df)$ represents a principal spline of space and time \citep[e.g.,][]{mardia1998kriged, sahu2005bayesian, fontanella2013functional} with $df$ bases. This strategy is preferred over a joint modeling of the exposure and the outcome because it avoids feedback from the outcome to the exposure that may lead to biased estimators of the exposure effect \citep[e.g.,][]{stephens2023causal}.

A Markov chain Monte Carlo (MCMC) scheme is implemented to simulate from the posterior distribution of $\boldTheta$. The algorithm proposed by \citet{chan2009efficient} is used to sample from the full conditionals of the spacetime-varying coefficients. It is an efficient simulation algorithm that allows to sample the coefficients for all instants of time in a single step. To avoid identifiability issues, all time-specific, area-specific and interaction effects are centered at zero during each MCMC iteration \citep[e.g.,][]{knorr2000bayesian,franco2022variance}. The latent process $\vartheta_{it}$ is sampled using Metropolis-Hastings steps. The remaining full conditional distributions are inverse-gamma distributions. Further details about the MCMC algorithm can be found in Section C of the Supplementary Material.

\section{Competing Approaches} \label{sec:tvc-competing-approaches}
This section briefly describes the competing approaches considered in the simulation study (in Section \ref{sec:tvc-simulations}) and in the real data application (in Section \ref{sec:tvc-application}). Similarly to SDGLMC, they are all based on the assumption that the outcome follows a Poisson distribution. The main differences lie in how the exposure effect is modeled and how unmeasured confounding is handled.

In each approach, the linear predictor is a function of exposure and measured confounders. To control for unmeasured confounding, additional covariates and basis functions are included in the linear predictor \citep[e.g.,][]{peng2006model,peng2008statistical}. The following models are fitted under the Bayesian paradigm: (i) the Null model, which does not account for unmeasured confounding; (ii) the GLMadj model, which adjusts for unmeasured confounding through a natural cubic spline of time \citep[e.g.,][]{peng2008statistical,zhou2021excess,masselot2022differential}; (iii) the JDZ model by \citet{janes2007trends}, which is similar to (ii) but further decomposes the exposure effect into two orthogonal parts; (iv) the Dummy model \citep[e.g.,][]{dai2014associations,klompmaker2023effects}, which allows for season-specific exposure effects; and (v) the Periodic model by \citet{peng2005seasonal} and \citet{chen2013seasonal}, which assumes that the exposure effect is a linear combination of trigonometric functions. Each model adopts a hierarchical structure, where spatial-specific coefficient vectors are assumed to be independently and identically distributed. Further details are provided in Section D of the Supplementary Material.

\section{Analysis of Artificial Data} \label{sec:tvc-simulations}
We conduct two simulation studies, inspired by the real data example presented in Section \ref{sec:tvc-motivating-example}, to validate the proposed approach.

\myParagraph{Simulation Study 1: Confounding Bias Mitigation}
The first study assesses the ability of SDGLMC to mitigate confounding bias.
We consider the $\ns=117$ contiguous HDs of Piemonte and Lombardia. We assume $\nt=600$ and $\pmeas = 0$ (i.e., there are no measured confounders). The exposure, $X_{it}$, and the unmeasured confounder, $Z_{it}$, are generated as correlated vector autoregressive processes of order $1$ [VAR($1$)] with a linear correlation coefficient of $0.5$. For both processes, the innovations are temporally independent but spatially structured. The spatial dependence is parameterized using a proper conditional autoregressive (PCAR) specification \citep{banerjee2014hierarchical,besag1974spatial,sahu2022bayesian}. The marginal spatio-temporal means of the exposure and the unmeasured confounder are generated from Fourier basis expansions. The correlation structure of $X_{it}$ is parameterized by spatial and temporal autocorrelation parameters, i.e., $\phi_x^S \in (0,1)$ and $\phi_x^T \in (0,1)$, representing the strength of the spatial and temporal dependence, respectively. Similarly, $\phi_z^S$ and $\phi_z^T$ represent the autocorrelation parameters of $Z_{it}$. Then, $Z_{it}$ is sampled conditionally on $X_{it}$. The parameters $\phi_x^S$, $\phi_x^T$, $\phi_z^S$, and $\phi_z^T$ are varied to define three scenarios that are summarized in Table \ref{tab:tvc-scenarios}.


The outcome variable is sampled from the Poisson distribution in Equation (\ref{eq:tvc-outcome}) using the following linear predictor:
\begin{equation*}
  \vartheta_{it} = o_{it} + \beta_{0} + \beta_{1it} X_{it} + Z_{it} + u_{it}, \quad u_{it} \iid \Norm(0, \tau_u^2)\,,
\end{equation*}
where $\tau_u^2= 10^{-3}$. Additional information on the data generation mechanism is provided in Section E of the Supplementary Material.

\begin{table}[!t]
  \renewcommand{\arraystretch}{1.3}
  \centering
  \caption{Simulation scenarios and corresponding definitions.}
  \label{tab:tvc-scenarios}
  \begin{tabular}{llcccc}
      \hline
      Scenario & Unmeasured confounder, $Z_{it}$ & $\phi_x^S$ & $\phi_z^S$ & $\phi_x^T$ & $\phi_z^T$ \\
      \hline
      S1 & Smoother than exposure in space and time & 0.2 & 0.98 & 0.2 & 0.98 \\ 
      S2 & Smoother than exposure only in space & 0.2 & 0.98 & 0.98 & 0.2 \\ 
      S3 & Smoother than exposure only in time & 0.98 & 0.2 & 0.2 & 0.98 \\
      \hline
  \end{tabular}
\end{table}

SDGLMC is compared to the Null model and to the GLMadj model here. The results from the other competing approaches can be found in Section E of the Supplementary Material. A natural cubic spline of time with $6$ degrees of freedom is used to account for unmeasured confounding in GLMadj. This choice follows from fitting several models with different numbers of bases and selecting the smallest number from which the estimates appear stabilized \citep{peng2006model}.

\begin{figure}[!htbp]
  \centering
  \includegraphics[width=\textwidth]{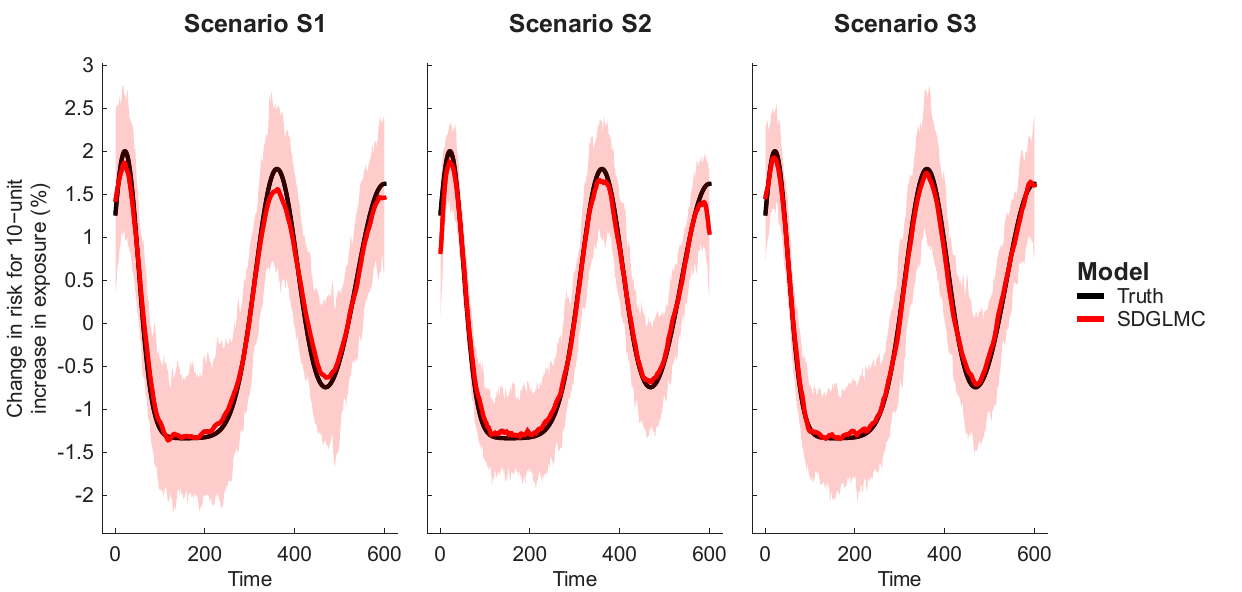}
  \caption[Simulation Study 1]{Simulation Study 1: temporal component of the exposure effect, expressed as percent change in risk associated with a 10-unit increase in the exposure, for each scenario. The red solid line represents the estimate (averaged over 100 replicates) obtained from SDGLMC, and the shaded area indicates the percentiles $2.5$ and $97.5$. The black solid line represents the true temporal component. Null and GLMadj models are not shown because they assume an exposure effect that is constant in time.}
  \label{fig:tvc-3scenarios}
\end{figure}

\begin{figure}[!htbp]
  \centering
  \includegraphics[width=14cm]{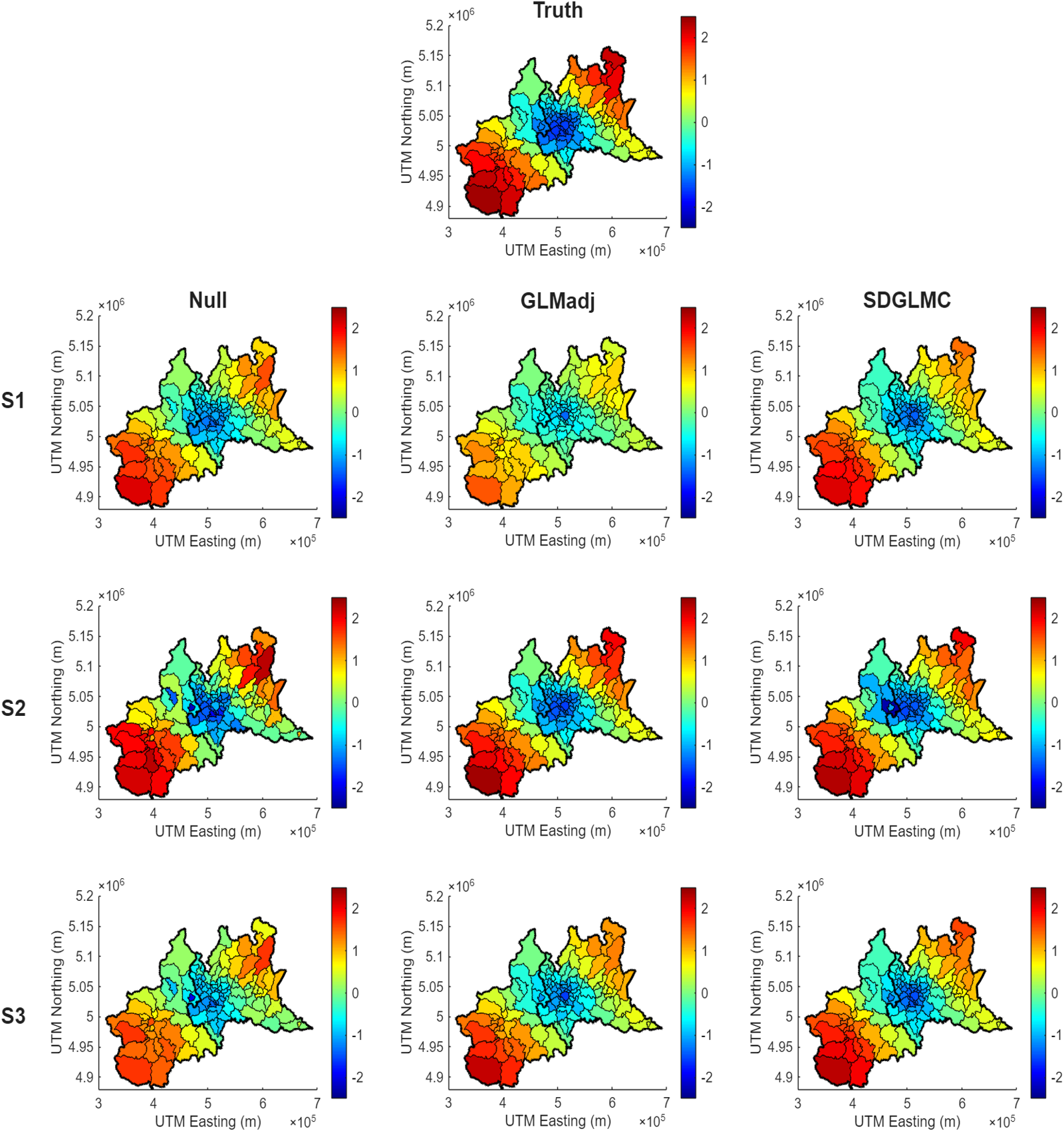}
  \caption[Simulation Study 1, space]{Simulation Study 1: spatial component of the exposure effect, expressed as percent change in risk associated with a 10-unit increase in the exposure. The top panel shows the true spatial component. The remaining panels show the posterior means (averaged over 100 replicates) under each scenario (rows) and model (columns).}
  \label{fig:tvc-bspace-3scenarios}
\end{figure}

The results are reported in Table \ref{tab:tvc-3scenarios-baseline-effect} and Figures \ref{fig:tvc-3scenarios} and \ref{fig:tvc-bspace-3scenarios} and are expressed as percent change in risk associated with a 10-unit increase in the exposure. For each scenario, the reported values summarize the distribution of the posterior means over the 100 replicates, namely their average and the $2.5$ and $97.5$ percentiles are taken into consideration.

Table \ref{tab:tvc-3scenarios-baseline-effect} compares the pooled estimates from the Null and GLMadj models to the baseline effect estimated from SDGLMC. Considering that the simulated value is $5\%$, the Null model exhibits a significant positive bias in all scenarios. GLMadj and SDGLMC effectively reduce (although without entirely removing) confounding bias in Scenario S1. In Scenario S2, the bias reduction is less pronounced. A possible explanation is that the unmeasured confounder varies at high frequencies in time, making it difficult to be adequately captured by time-varying intercepts (or natural splines of time in the case of GLMadj) that only account for gradual changes. The difference $X_{it} - \widehat{X}_{it}$ is approximately as rough as $Z_{it}$ in time. Also, as all these variables are correlated, part of the variability of the unmeasured confounder is explained by the local exposure effect. Performance under Scenario S3 is very similar to Scenario S1, suggesting that the lack of smoothness in space of the unmeasured confounder does not seem to alter the performance in terms of confounding bias of the proposed approach.

\begin{table}[!hbtp]
  \renewcommand{\arraystretch}{1.3}
  \centering
  \caption{Simulation Study 1: baseline component of the exposure effect under each scenario, expressed as percent change in risk associated with a 10-unit increase in the exposure. Values represent the average posterior means across 100 replicates, with $2.5$ and $97.5$ percentiles in brackets. The true simulated effect is $5\%$.}
  \label{tab:tvc-3scenarios-baseline-effect}
  \begin{tabular}{lccc}
      \hline
      Scenario & Null & GLMadj & SDGLMC\\
      \hline
      S1 & 10.427 [9.507, 11.310] & 5.189 [4.233, 5.872] & 5.282 [5.020, 5.568] \\ 
      S2 & 8.394 [8.118, 8.713] & 7.061 [6.809, 7.528] & 7.145 [6.849, 7.637] \\ 
      S3 & 10.028 [9.406, 10.688] & 5.462 [5.120, 6.286] & 5.568 [5.217, 6.320] \\
      \hline
  \end{tabular}
\end{table}

Figure \ref{fig:tvc-3scenarios} depicts the temporal component of the exposure effect. For each scenario, the black solid line represents the true effect. The estimate obtained from SDGLMC is shown in red, with the solid line indicating its average over 100 replicates, and the shaded area indicating the percentiles $2.5$ and $97.5$. The proposed approach successfully recovers the simulated temporal pattern in all scenarios, as opposed to the Null and GLMadj models, which do not account for time-varying effects.

Figure \ref{fig:tvc-bspace-3scenarios} shows the spatial component of the exposure effect for each scenario. The (averaged) posterior means obtained from SDGLMC are compared to the (averaged) posterior means of the differences between area-specific and pooled estimates obtained from the competing approaches. All methods generally recover the simulated spatial pattern, with lower values in the middle of the study area. SDGLMC provides accurate estimates in all scenarios, while the Null and GLMadj models tend to underestimate the spatial effect in some districts.

The above results are also confirmed by Tables E.4--E.6 in the Supplementary Material, where the average mean squared error ($MSE_{avg}$), the maximum absolute bias ($MAB$), and the average mean absolute error ($MAE_{avg}$) are reported for each model and each scenario (see Section E.2 in the Supplementary Material for definitions). 
We also report these metrics for each component of the exposure effect to assess their contribution to the overall error. In general, we note that SDGLMC outperforms the competing models in all scenarios.

\myParagraph{Simulation Study 2: Recovery of the Local Exposure Effect ($f'$) and Model Reparameterization}
The second simulation study is fully described in Section F of the Supplementary Material. A first objective is to assess whether the proposed approach is able to recover the first derivative $f'(\cdot; \boldgamma_f)$ when the data are generated using Equations (\ref{eq:tvc-outcome})--(\ref{eq:tvc-linear-predictor}). We show that SDGLMC is able to capture the dynamics of the exposure effect, although a residual bias remains due to confounding.

Another objective of the second simulation study is to provide practical motivation for the decomposition of the exposure effect into local-scale and large-scale components in Equation (\ref{eq:tvc-taylor}). We compare SDGLMC with a reparameterized model that does not remove the large-scale spatio-temporal trend from the exposure. As expected from the discussion in Section \ref{sec:tvc-model}, the results suggest that the proposed approach is more effective in reducing the confounding bias and always has lower $MSE_{avg}$, $MAB$ and $MAE_{avg}$ values (see Table F.7 in the Supplementary Material).

\section{Application to the Italian Mortality Data} \label{sec:tvc-application}

We apply SDGLMC to the Italian mortality dataset (see Section \ref{sec:tvc-motivating-example}) and compare its results to those obtained from the competing approaches. The large-scale trend, $\widehat{X}_{it}$, is estimated using a principal spline with $5$ spatial basis dimensions and $20$ temporal basis dimensions. For each HD and instant of time, $\MM_{it}$ includes bases from natural cubic splines of daily temperature and daily relative humidity (with 7 and 4 degrees of freedom, respectively), along with indicator variables for weekdays, holidays and lockdown periods (thus, $\pmeas=19$; see also Section A of the Supplementary Material).

\myParagraph{Types of Space-Time Interaction Component, $\delta_{0it}^*$}
We can consider several distributional specifications for the space-time interaction component of the intercept, $\delta_{0it}^*$, depending on the correlation structure we want to impose. \citet{knorr2000bayesian} proposes four types of models, each defining a different space-time interaction term characterized by distinct prior assumptions. These models are reviewed by \citet{franco2022variance} and \citet{orozco2023big} and summarized in Table \ref{tab:tvc-priors-knorr-held}. Type T1 assumes unstructured space-time variability and is obtained by setting $\delta_{0it}^* = 0$ for all $i$ and $t$ as an unstructured error is already present in Equation \eqref{eq:tvc-obs_proposal}. Type T2 assumes independent temporal trends for each HD, with a common variance for the innovations. Type T3 assumes independent spatial trends for each time $t$, with a common conditional variance. Type T4 assumes complete dependence, i.e., the temporal trends are different across HDs but are more likely to be similar for neighboring HDs.

The prior assumptions discussed in \citet{knorr2000bayesian} can be further relaxed. In fact, during the pandemic, some HDs in Lombardia --- more specifically in the province of Bergamo --- experienced a much greater increase in deaths than any other HD. To capture such heterogeneity, we introduce a fifth interaction type, T5, which assumes independent space-time interactions across HDs following random walk processes with area-specific evolution variances. Thus, interaction T5 captures temporal trends that differ considerably among HDs but do not have any spatial structure.

\bgroup
\def\arraystretch{1.4}
\begin{table}[!htbp]
  \centering
  \caption{Space-time interaction types considered for $\delta_{0it}^*$. DIC values are reported for model selection; the lowest value is in italics.}
  \label{tab:tvc-priors-knorr-held}
  \begin{tabular}{l l l l}
    \hline
    Type & Space-Time structure & Distributional Assumption & DIC\\
    \hline
    T1 & Absent & $\delta_{0it}^*=0$ & 814038 \\
    T2 & \makecell[l]{Independent RW,\\ same variance} & $\delta_{0it}^* \sim \Norm (\delta_{0i,t-1}^*, \sigma_{w0}^{* 2})$& 807711\\
    T3 & Independent ICAR & $ \bolddelta_{0t}^* \sim ICAR(\sigma_{w0}^{* 2}) $& 811293 \\
    T4 & Complete dependence & $ \bolddelta_{0t}^* - \bolddelta_{0,t-1}^* \sim ICAR(\sigma_{w0}^{* 2}) $ & 811923\\ 
    T5 & \makecell[l]{Independent RW,\\ different variances} & $\delta_{0it}^* \sim \Norm (\delta_{0i,t-1}^*, \sigma_{w0i}^{* 2})$ & \textit{803861} \\
    \hline
  \end{tabular}
\end{table}
\egroup

\myParagraph{Model Fitting and Presentation of Results}
The MCMC algorithm described in Section \ref{sec:tvc-inference} is run for 100,000 iterations, and posterior inference is based on the final 80,000 draws. The model's chains are monitored for convergence visually through trace plots and using the R-statistic of \citet{gelman1996inference} on two chains that are started from different initial points simultaneously.

After fitting a model for each of the five types of space-time interaction components, the one with the lowest deviance information criterion \citep[DIC;][]{spiegelhalter2002bayesian} is selected. As shown in the last column of Table \ref{tab:tvc-priors-knorr-held}, the model with interaction type T5 has the smallest DIC value, probably because it is able to capture highly specific spatio-temporal trends---such as those observed in the Bergamo area during the pandemic. Thus, the results of the model with interaction type T5 are discussed in detail.

To assess the fit of the posterior distribution of SDGLMC, we compare the observed data to the posterior predictive distribution. For each HD and time point, we compute a Bayesian $p$-value based on the test quantity $\Pr\bigl(Y_{it}> Y^{rep}_{it} \bigr) + \frac{1}{2} \Pr\bigl(Y_{it} = Y^{rep}_{it} \bigr)$, where $Y^{rep}_{it}$ is generated from the posterior predictive distribution at each MCMC iteration. Values near $0.5$ suggest that the model provides an adequate fit to the observed data. Conversely, values approaching the extremes of the unit interval (i.e., $<0.1$ or $>0.9$) indicate poor fit \citep{gelman2013bayesian}. The estimated mean $p$-value is approximately $0.5$, supporting the adequacy of the model.

\begin{figure}[!htbp]
  \centering
  \includegraphics[width=12cm]{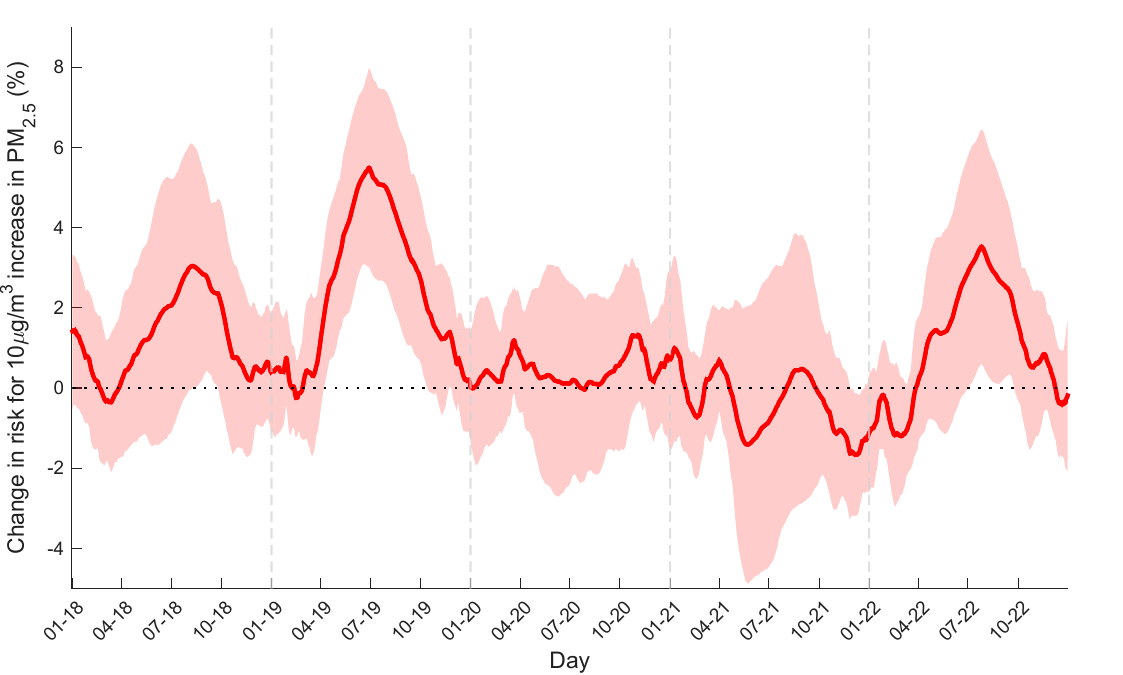}
  \caption[Estimated time-varying percent change in risk]{Posterior summary (solid line: mean, shaded area: point-wise limits of 95\% posterior credible interval) of the space-averaged local exposure effect, expressed as percent change in mortality risk for a $10\ \mu g/m^3$ increase in \pmtwopfive\ concentrations. Vertical dashed lines mark calendar years.}
  \label{fig:tvc-application-beta1t-soloDGLM}
\end{figure}

Figure \ref{fig:tvc-application-beta1t-soloDGLM} depicts the estimated space-averaged local exposure effect, expressed as the percent change in risk associated with a $10\ \mu g/m^3$ increase in the \pmtwopfive\ concentrations. In other words, the quantity $100(\exp \{10 (\overline{\delta}_1 + \delta_{1t}) \}-1)$ is computed for each MCMC iteration, which represents the average effect across all HDs at a given time $t$. The red solid line represents the posterior mean, and the shaded area denotes the 95\% posterior credible interval (CI). The exposure effect is not constant in time as no horizontal straight line can be drawn within the CI. Instead, it exhibits distinct peaks across different years, and three phases can be identified: 2018-19, 2020-21 and 2022.

During the first two years, the local exposure effect exhibits a seasonal pattern, with the peaks occurring during the summer months. The estimated association is bounded away from zero between August and September 2018 and from April to October 2019, with the highest change in mortality risk of $5.50\%$ (95\% CI: $[3.03\%, 7.98\%]$) observed on June 29, 2019. In the second phase (2020-21), the CI generally includes zero, except for a few days toward the end of the period. Finally, in the third phase (2022), the shape of the local exposure effect resembles that of the first phase, remaining bounded away from zero between July and September, with the highest change in mortality risk of $3.53\%$ (95\% CI: $[0.61\%, 6.44\%]$) occurring on July 26. 

\begin{figure}[!tbp]
  \centering
  \includegraphics[width=13cm]{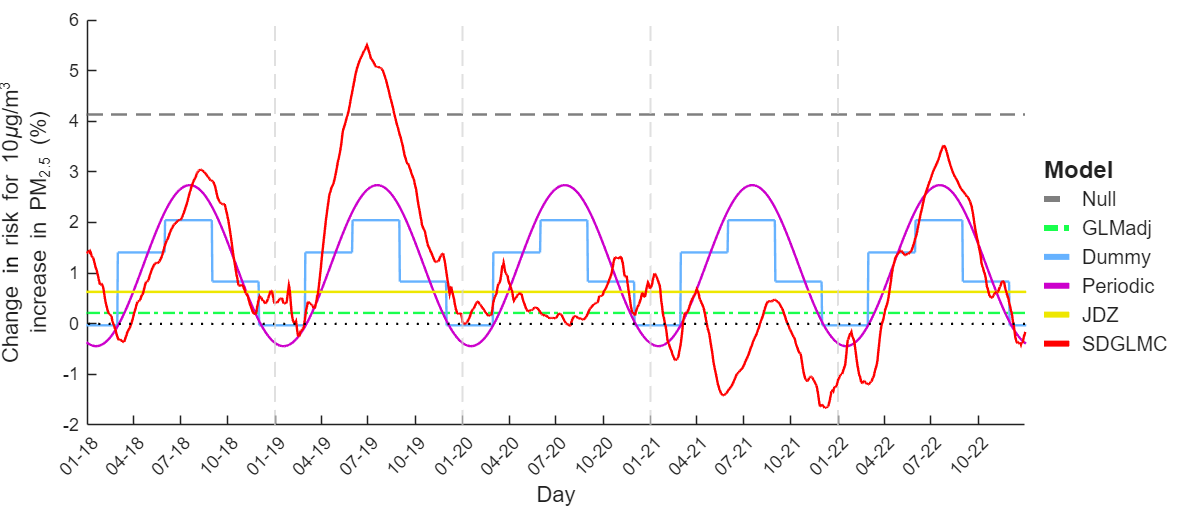}
  \caption[Estimated percent change in risk under various approaches]{Estimated percent change in risk associated with a $10\ \mu g/m^3$ increase in \pmtwopfive\ concentrations: comparison of SDGLMC with the approaches described in Section \ref{sec:tvc-competing-approaches}. Vertical dashed lines mark calendar years. The horizontal dotted line indicates no effect.}
  \label{fig:tvc-competing-models-beta1t}
\end{figure}

\myParagraph{Comparison with Competing Approaches}
For each approach introduced in Section \ref{sec:tvc-competing-approaches}, Figure \ref{fig:tvc-competing-models-beta1t} shows the pooled estimated percent change in risk associated with a $10\ \mu g/m^3$ increase in \pmtwopfive\ levels. For comparison, the posterior mean from SDGLMC is also shown in red. The Null and GLMadj models yield the highest ($3.92 \%$) and the lowest ($0.21\%$) estimates on average, respectively. The Null, GLMadj and JDZ models assume that the association of interest is constant over time; however, we argue that this assumption is unlikely to hold (see Section \ref{sec:tvc-introduction}).

Among the competing approaches, only the Dummy and Periodic models are able to capture time-varying strengths of the association. They seem to agree with SDGLMC on the temporal evolution of the effect, with peaks in summer and troughs in winter. Although the effect estimates change within the year, they are constant across years. This assumption is limiting, as the pandemic outbreak may have altered the strength of the association. Note also that the Dummy model allows for abrupt changes in the exposure effect; however, these are unlikely in practice as pointed out by \citet{bhaskaran2013time} and \citet{peng2005seasonal}. Moreover, the resulting estimates may be too sensitive to the choice of the boundaries of each season, and a definition based on the calendar seasons may result suboptimal.

\begin{figure}[!htbp]
  \centering
  \includegraphics[width=14cm]{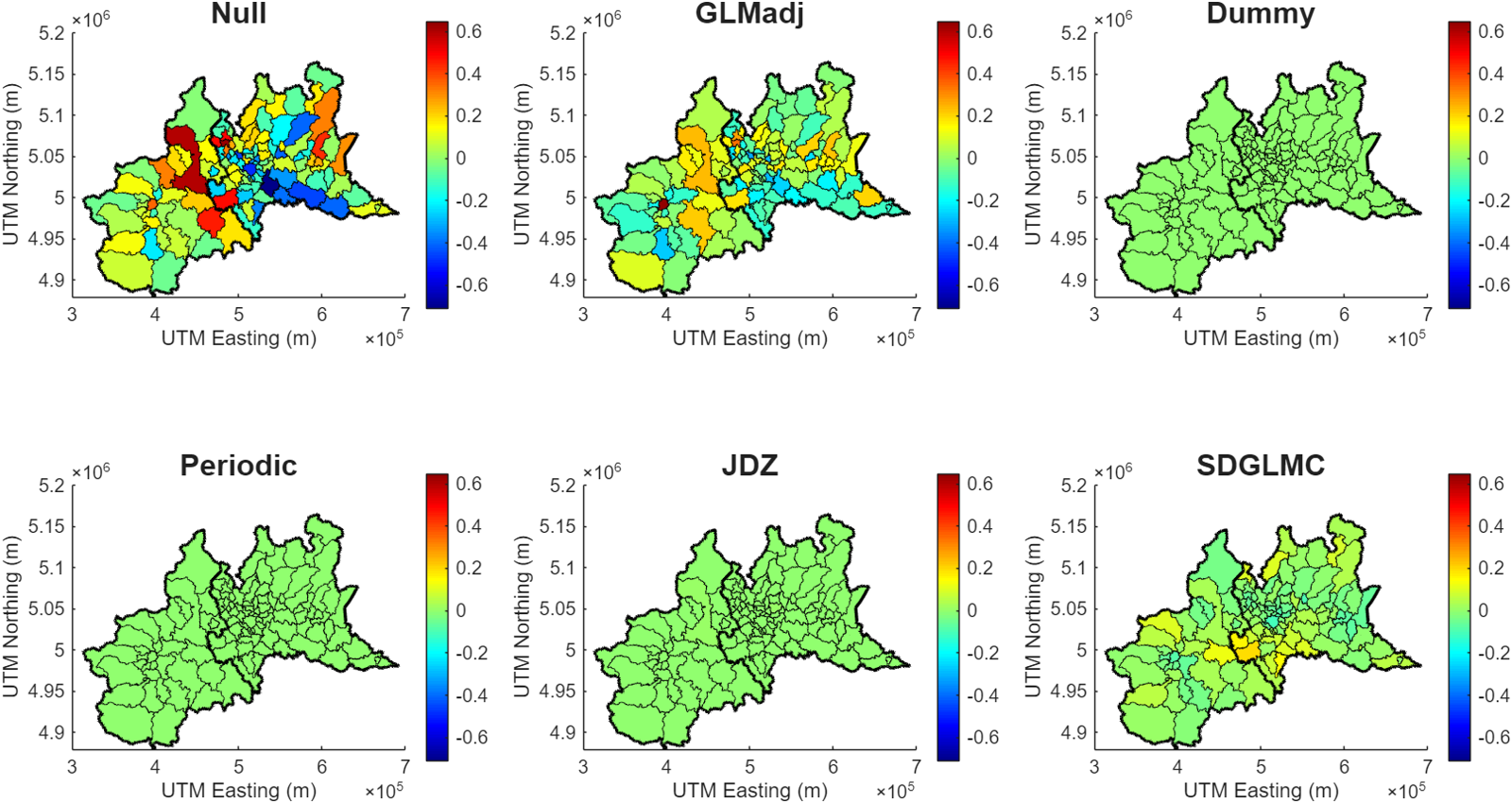}
  \caption{Posterior means of the spatial component of the percent change in risk associated with a $10\ \mu g/m^3$ increase in \pmtwopfive\ concentrations under each approach.}
  \label{fig:tvc-competing-models-beta1t-spatialeffect}
\end{figure}

Regarding spatial variation, we compare the differences between area-specific and pooled estimates under each competing approach with the spatial component of the local exposure effect ($\Tilde{\bolddelta}_{1}$). Figure \ref{fig:tvc-competing-models-beta1t-spatialeffect} shows the posterior means of these quantities, expressed as percent change in risk associated with a $10\ \mu g/m^3$ increase in \pmtwopfive\ concentrations. The spatial variability of the exposure effect approaches zero for all models except the Null and GLMadj. However, when considering the associated uncertainty, there is no evidence of spatial variation in the exposure effect for any of the models. This finding may be related to the relatively small size of the study region and the similarity between Piemonte and Lombardia (e.g., in terms of socio-economic indicators).

\myParagraph{Comparison with a Restricted Model}
In Section G of the Supplementary Material, we investigate the consequences of removing the measured confounders (i.e., temperature, relative humidity, and indicator variables for weekdays, national holidays, and lockdown periods) from the model specification. In other words, we impose the restriction that the elements of $\boldalpha$ in Equation (\ref{eq:tvc-latentBETA}) are set to zero. The findings are insightful: the estimated exposure effect greatly overlaps the estimate obtained when the measured confounders are included, with differences occurring only during the summer months. Moreover, it suggests that SDGLMC can provide accurate estimates even in complex scenarios, and that the confounding effect may be time-varying and due to heat waves.

\section{Discussion} \label{sec:tvc-conclusions}
Motivated by a real case-study, this paper proposes the \textit{spatial dynamic generalized linear model for confounding (SDGLMC)} to address two key issues in observational studies that may distort inference and interpretation of the exposure--outcome relationship, namely misspecification bias and confounding bias. Thanks to the first-order Taylor's expansion, the exposure effect is distinguished into large-scale and small-scale components. To recover the small-scale spatio-temporal association, the outcome is regressed on a detrended exposure. This approach is beneficial because detrending improves estimation stability and helps reduce confounding bias \citep{janes2007trends,greven2011approach}. Furthermore, the inclusion of spacetime-varying coefficients allows SDGLMC to overcome both types of bias. As shown in Section \ref{sec:tvc-simulations}, SDGLMC is able to almost recover the true functional form of the exposure effect. Thus, it should be preferred over models that assume a constant association.

A similar decomposition of the exposure effect is proposed by \citet{janes2007trends} and \citet{dupont2022spatial+}; however, they assume a constant effect. Another connection exists between SDGLMC and the work by \citet{dupont2022spatial+}, as the smooth function of spatial coordinates in their model and the spatio-temporal intercept in SDGLMC serve a similar role.

The estimated local exposure effect exhibits a seasonal pattern, with a higher percent change in risk observed during the summer months, when the \pmtwopfive\ concentrations are generally lower. This behavior indicates that a threshold below which the exposure effect is null may not exist \citep[see, among others,][]{brunekreef2021mortality,cork2023methods,dominici2022assessing,renzi2021acute}. Additionally, the behavior suggests that the unmeasured confounders could have an interaction effect with the exposure (affecting $\beta_{1it}$). The seasonal pattern might be due to potential interactions between \pmtwopfive\ concentrations and air temperature. As noted in Section \ref{sec:tvc-introduction}, increased ventilation and physiological susceptibility to air pollution may explain the higher percent change in risk during summer. Further, interactions with other variables, such as other pollutants, are also possible \citep[e.g.,][]{liu2023interactive, ji2022no2, siddika2019synergistic}. Similar findings have been reported in other studies \citep{nawrot2007stronger,peng2005seasonal, peng2006model, roberts2004interactions, stafoggia2008does, sun2015temperature, tian2018effects}. Although the higher exposure effect in periods with lower \pmtwopfive\ concentrations may be counterintuitive, the interaction mechanisms described above offer a plausible explanation.

Fine particulate matter is a complex mixture of organic (e.g., black carbon and organic carbon) and inorganic compounds (e.g., sulfate, nitrate, and ammonium), and the relative proportion of these components may vary in time \citep{franklin2008role, masselot2022differential, vinson2020seasonal}. Their precursors are emitted from both natural (e.g., oceans and volcanoes) and anthropogenic sources, such as fossil fuel combustion, industrial processes, and residential heating. Their transportation and transformation into \pmtwopfive\ in the atmosphere are mainly influenced by meteorological conditions \citep{hao2023national,jing2020effects,kelly2012size}. Toxicological studies indicate that the PM constituents have different effects on human health \citep[e.g.,][]{kelly2012size}, and some authors suggest that ammonium is probably among the most harmful components \citep{masselot2022differential,hao2023national}. The relative proportions of the \pmtwopfive\ components have not been analyzed in this study, so they could be the subject of future research. Notably, \citet{hao2023national} identify a non-linear long-term association between some \pmtwopfive\ constituents and mortality.

An exception to the seasonal behavior is observed in 2020 and 2021, when the exposure effect is not bounded away from zero. This is likely attributable to lockdown measures --- such as stay-at-home orders and widespread mask usage --- which may have reduced exposure to air pollution. A similar result is found in the absence of measured confounders (see Figure G.1 in the Supplementary Material), and is also supported by recent studies \citep[e.g.,][]{hadei2021effect, lopez2021air}. It is worth recalling that none of the competing approaches in Section \ref{sec:tvc-competing-approaches} are able to capture the deviation from the seasonal behavior.

The literature reveals a lack of consensus regarding the association between air pollution and adverse COVID-19 health outcomes in the general population \citep[e.g.,][]{brunekreef2021air,walton2021investigating}. More recently, \citet{bhaskar2023literature} review previous studies and report that the majority suggests a positive association. In contrast, our findings indicate that the local exposure effect in 2020 and 2021 is generally not bounded away from zero (except for a few days). Several factors may explain this result. First, our model uses all-cause mortality among the elderly population as the outcome, making it not directly comparable to COVID-19-specific studies. Second, the issuance of the lockdown measures by the Italian government, along with the mandatory use of facial masks, not only effectively curbed the transmission of the SARS-CoV-2 virus (and of other viruses), but also potentially reduced exposure to air pollution. During lockdown periods, people were required to stay at home, resulting in reduced exposure to outdoor air pollutants. Ventilation was also reduced because lockdowns occurred mainly during winter months. Moreover, the use of facial masks may have reduced the inhalation of airborne hazardous particles \citep{kodros2021quantifying,zhou2018assessment}.

Although lockdown periods were not issued in the second half of 2021, the local exposure effect remains not bounded away from zero. This could be due to continued preventing behaviors such as mask-wearing and hand sanitization. Another possible explanation is \textit{mortality displacement} (or \textit{harvesting}), where the most vulnerable people may have died earlier than expected in the preceding years due to the SARS-CoV-2 virus in 2020-2021, resulting in an apparently null (or even protective) effect of the exposure afterwards \citep{armstrong2017longer,bhaskaran2013time,zanobetti2008mortality,zanobetti2000generalized}.

During the third phase, people started to return to their normal activities, resulting in increased exposure to air pollution. Consequently, the local exposure effect has a shape similar to that in the first phase.

Separate analyses conducted for each year suggest patterns that are qualitatively consistent with our findings. Due to the limited temporal scope, this analysis is restricted to the GLMadj and JDZ models. Table \ref{tab:tvc-appl-others} shows that both models indicate a non-significant exposure effect during the second phase.

\begin{table}[!ht]
  \centering
  \caption{Year-specific estimates of the exposure effect from GLMadj and JDZ models, expressed as percent change in risk associated with a $10\ \mu g/m^3$ increase in the \pmtwopfive\ concentrations. Posterior means are shown for each year from 2018 to 2022, with 95\% CIs reported in brackets.}
  \begin{tabular}{lccccc}
  \hline
      ~ & 2018 & 2019 & 2020 & 2021 & 2022 \\ \hline
      GLMadj & 0.75 [0.06, 1.45] & 0.81 [0.22, 1.39] & 0.14 [-0.43, 0.69] & -0.03 [-0.69, 0.68] & 0.67 [0.01, 1.31] \\ 
      JDZ & 0.82 [0.13, 1.50] & 0.85 [0.25, 1.43] & 0.26 [-0.30, 0.81] & -0.03 [-0.69, 0.68] & 0.58 [-0.07, 1.22] \\ \hline
  \end{tabular}
  \label{tab:tvc-appl-others}
\end{table}

Unmeasured confounding may induce spatial and spatio-temporal correlation in the exposure effect and in the residuals. Thus, SDGLMC includes spatial and spatio-temporal components in the intercept that are structured according to the adjacency of the areal units. Besides, SDGLMC accommodates potential spatial variation in the local exposure effect, which may reveal differences related to urbanicity and morphological characteristics of the study area. However, in the application to the Italian mortality data, the local exposure effect appears homogeneous across space. This is likely due to the relatively small size of the study area and the similar socio-economic and environmental profiles of Piemonte and Lombardia.

A reviewer pointed out that if the effect modification is solely due to measured confounders, it can be more straightforwardly examined by including interaction terms between \pmtwopfive\ and the confounders in the linear predictor. We explore this possibility in Section G.2 of the Supplementary Material. However, since the estimated effect is substantially different from that shown in Figure \ref{fig:tvc-application-beta1t-soloDGLM} --- particularly during the summer months and the pandemic --- we conclude that the effect modification is likely due to both measured and unmeasured confounders.

\citet{janes2007trends} propose a method for testing the presence of unmeasured confounding when the exposure effect is partitioned into two orthogonal components. According to their approach, if the overall and local exposure effects are equal, it is likely that confounding bias is absent. While this testing procedure is theoretically feasible in our case, its practical implementation is challenging as it requires specific assumptions about the functional form of $f(\cdot; \boldgamma_f)$.

It is arguable that, even if SDGLMC could recover the true exposure effect $f(\cdot; \cdot)$, its exact functional form remains unknown. In other words, the complete set of interaction terms cannot be determined. However, we believe this limitation is not unique to the proposed model. In fact, the list of interaction terms is generally unknown when fitting any regression model, and it is up to the researcher to choose the most appropriate form for $f(\cdot; \cdot)$. This choice inherently involves a degree of subjectivity and uncertainty. By contrast, the proposed approach offers greater flexibility and should be able to capture the effects of lockdowns as well as interaction effects between exposure and other variables, such as other pollutants, thereby mitigating inferential issues due to misspecification.

Further research could explore the causal interpretation of the estimated exposure effect, which would require additional assumptions \citep[e.g.,][]{hernan2020causal}. However, it is important to emphasize that the analytical approach adopted here should not be considered as less credible than causal inference methods \citep{dominici2017best}.

In conclusion, SDGLMC is a flexible and effective approach to estimate the exposure--outcome association in the presence of spacetime-varying effects and unmeasured spatio-temporal confounding. The model is able to capture the complex dynamics of the exposure effect while controlling for confounding. The proposed approach is a valuable tool for researchers interested in the small-scale, spatio-temporal health effects of air pollution and can be extended to other fields where the exposure--outcome association is of interest.

\section*{Acknowledgements}

The authors are grateful to the Associate Editor and two anonymous reviewers for their constructive comments that helped improve the manuscript.
\citet{hersbach2023era5} was downloaded from the \citet{copernicus2023era5}. \citet{CAMSdataset} was downloaded from the \citet{copernicus2023cams}. The results contain modified C3S and CAMS information. Neither the European Commission nor ECMWF is responsible for any use that may be made of the Copernicus information or data it contains. The research of C.Z., P.V., and L.I. is funded by the European Union - NextGenerationEU. C.Z.: PRIN PNRR 2022 (Bando 2023) \textit{SLIDE -  Stochastic Modeling of Compound Events} (P2022KZJTZ - CUP D53D23018920001). P.V. and L.I.: research project PRIN2022 \textit{CoEnv - Complex Environmental Data and Modeling} (2022E3RY23 - CUP D53D23011080006). A.M.S. is grateful for financial support from the Natural Sciences and Engineering Research Council (NSERC) of Canada (Discovery Grant RGPIN-2024-04312).\vspace*{-8pt}

\section*{Code Availability}

The code associated to this paper is available at \url{https://github.com/czaccard/SDGLMC}.\vspace*{-8pt}



\bibliographystyle{abbrvnat} 
\bibliography{bibliography}

\clearpage

\section*{{\huge Supplementary Materials}}
\appendix

\setcounter{figure}{0}
\setcounter{table}{0}
\renewcommand*{\thesection}{\Alph{section}}
\renewcommand*{\thefigure}{%
  \Alph{section}.\arabic{figure}%
}
\renewcommand*{\thetable}{%
  \Alph{section}.\arabic{table}%
}

\section{Data Collection} \label{sec:tvc-data-collection}

We gather daily data to analyze the short-term association between \pmtwopfive\ and all-cause mortality in Piemonte and Lombardia regions for people aged 65 or more. The data span the period between 2018 and 2022, and are collected at the health district (HD) level. 

\begin{itemize}
  \item \textbf{Mortality data.} All-cause mortality data are collected from the Italian National Institute of Statistics \citep{istat_decessi_comuni} for each municipality in the two regions, and are then spatially aggregated to the HD level, which represents the areal unit. The spatial domain comprises $\ns=117$ HDs where $84$ belong to the Lombardia region and $33$ to the Piemonte region. The municipality of Campione d'Italia is excluded from the analysis since it has a small area and the time series contains many zeroes. Since the data refer to five consecutive years, the temporal dimension is $\nt=1826$.
  
  For $i=1, \dots, \ns$ and $t=1, \dots, \nt$, the standardized mortality ratio (SMR) is calculated as the ratio between the observed number of deaths, $Y_{it}$, and the expected number of deaths, $E_{it}$. The latter is obtained by indirect standardization \citep{banerjee2014hierarchical,sahu2022bayesian} as follows. The resident population aged 65 or more represents the population at risk and is obtained from the Italian National Institute of Statistics for each year \citep{istat_popolazione}. The population is stratified by age class (65--69, 70--74, 75--79, 80--84, 85--89, 90--94, or 95+ years) and gender (male or female). For each stratum $k = 1, \dots, K$, the mortality rate for the northwest Italy is retrieved from the Italian National Institute of Statistics for each year \citep{istat_mortality_table}. The expected counts are thus computed as follows:
  \begin{equation*}
    E_{it} = \sum_{k=1}^{K} Pop_{itk} r_{tk}\,,
  \end{equation*}
  where $Pop_{itk}$ and $r_{tk}$ are the population at risk and the mortality rate in the $k$-th stratum, respectively.

  \item \textbf{Exposure data.} Data on \pmtwopfive\ concentrations (expressed in $\mu g/m^3$) are collected from the Copernicus Atmosphere Monitoring Service's (CAMS) Atmosphere Data Store \citep{CAMSdataset,copernicus2023cams}. Hourly data are available at a spatial resolution of $0.1^{\circ} \times 0.1^{\circ}$, so we use zonal statistics to aggregate the estimates to the HD level, and then average across time to obtain daily data. 

  \item \textbf{Potential confounders.} Daily temperature, daily relative humidity, indicator variables for the days of the week, for national holidays and for lockdown periods are treated as potential measured confounders. We gather satellite measurements of air temperature and dew-point temperature (both expressed in Kelvin degrees) from the ERA5 reanalyses dataset available at the Copernicus Climate Change Service's (C3S) Climate Data Store \citep{copernicus2023era5,hersbach2023era5}. Relative humidity ($RH$) is calculated from air temperature ($Temp$) dew-point temperature ($DT$) using the August-Roche-Magnus approximation formula \citep{alduchov1996improved}:
  \begin{equation*}
    RH = 100 \times \exp \left(\frac{17.625 \times DT}{243.04+DT}-\frac{17.625 \times Temp}{243.04+Temp}\right)\,.
  \end{equation*}

  These hourly data are spatially and temporally averaged, similarly to the exposure. 

  The lockdowns are defined as follows: the first period is from March 9, 2020, to May 3, 2020, the second one is from December 21, 2020, to January 15, 2021, and the third one from March 6, 2021, to April 25, 2021.

\end{itemize}

\section{Descriptive Statistics} \label{sec:tvc-descriptive-statistics}

Descriptive statistics on the SMR data in all HDs are reported in Table \ref{tab:tvc-mortality-regions}. The average SMR is $1.28$ and $1.18$ in Lombardia and Piemonte, respectively, and the standard deviation is slightly higher in Lombardia ($0.34$) than in Piemonte ($0.26$).

\begin{table}[!htbp]
  \centering
  \caption{Descriptive statistics of daily SMR across all HDs in 2018--2022. Q1, Q2, and Q3 stand for first, second, and third quartiles, respectively.}
  \label{tab:tvc-mortality-regions}
  \begin{tabular}{cccccccc}
  \hline
  Region & Mean & SD & Min & Q1 & Q2 & Q3 & Max\\
  \hline
  Lombardia & 1.28 & 0.29 & 0.75 & 1.14 & 1.25 & 1.39 & 3.52\\
  Piemonte & 1.18 & 0.22 & 0.68 & 1.03 & 1.14 & 1.29 & 2.10\\
  \textbf{Overall} & 1.25 & 0.26 & 0.75 & 1.12 & 1.22 & 1.37 & 3.09\\
  \hline
  \end{tabular}
\end{table}

Descriptive statistics on \pmtwopfive\ levels in all HDs during the whole study period are reported in Table \ref{tab:tvc-covariates-regions}. The average daily \pmtwopfive\ concentration is $18.21 \, \mu g/m^3$ (standard deviation = $10.36$) over the entire spatial domain, but Lombardia is more polluted than Piemonte. Table \ref{tab:tvc-covariates-regions} also reports descriptive statistics of temperature and relative humidity. The average daily temperature is $284.91 \, K$ ($7.60$), while the average daily relative humidity is $91.76 \%$ ($3.59$). These are very similar to the region-specific values (not shown).

\begin{table}[!htbp]
  \centering
  \caption{Descriptive statistics of daily \pmtwopfive\ concentrations, temperature, and relative humidity across all HDs in 2018--2022. Q1, Q2, and Q3 stand for first, second, and third quartiles, respectively.}
  \label{tab:tvc-covariates-regions}
  \begin{tabular}{cccccccc}
  \hline
  Variable & Mean & SD & Min & Q1 & Q2 & Q3 & Max\\
  \hline
  \pmtwopfive\ ($\mu g/m^3$) & 18.21 & 10.36 & 2.32 & 10.49 & 15.20 & 24.12 & 63.07\\
  Temperature ($K$) & 284.91 & 7.60 & 265.15 & 278.17 & 284.68 & 292.02 & 301.59\\
  Relative humidity (\%) & 91.76 & 3.59 & 79.55 & 89.43 & 92.12 & 94.54 & 98.43\\
  \hline
  \end{tabular}
\end{table}

Finally, Table \ref{tab:tvc-annual-means} reports the annual death counts and annual averages of the daily SMR and daily \pmtwopfive\ for each administrative region.

\begin{table}[htbp]
  \centering
  \caption{Annual mortality counts, and annual averages of daily SMR and daily \pmtwopfive\ concentrations for Piemonte and Lombardia (2018--2022).}
  \label{tab:tvc-annual-means}
  \begin{tabular}{ccccccc}
    \hline
    \multirow{2}{*}{~} & \multirow{2}{*}{~} & \multicolumn{5}{c}{Year} \\
    \cline{3-7}
    Variable & Region & 2018 & 2019 & 2020 & 2021 & 2022 \\
    \hline
    \multirow{3}{*}{Mortality Counts} & Lombardia & 89742 & 90613 & 123586 & 97365 & 101069 \\
    & Piemonte & 48739 & 48479 & 60364 & 51202 & 53581 \\
    & \textbf{Overall} & 138481 & 139092 & 183950 & 148567 & 154650 \\
    \hline
    \multirow{3}{*}{SMR} & Lombardia & 1.32 & 1.32 & 1.23 & 1.29 & 1.26 \\
    & Piemonte & 1.20 & 1.20 & 1.16 & 1.18 & 1.17 \\
    & \textbf{Overall} & 1.29 & 1.29 & 1.21 & 1.26 & 1.24 \\
    \hline
    \multirow{3}{*}{PM$_{2.5}$ ($\mu g/m^3$)} & Lombardia & 18.56 & 20.41 & 18.36 & 19.42 & 20.30 \\
    & Piemonte & 14.00 & 16.47 & 14.33 & 15.17 & 15.91 \\
    & \textbf{Overall} & 17.27 & 19.30 & 17.22 & 18.22 & 19.06 \\
    \hline
  \end{tabular}
\end{table}

\section{Posterior Full Conditional Distributions for the Proposed DGLM} \label{sec:tvc-post-dist}

We provide the computational details to make inference on the model proposed in Section 3 and following the discussion in Section 4 of the main manuscript. In particular, stacking Equation (6) of the main text over the $\nt$ time instants, we have \citep[see also][]{chan2009efficient}:

\begin{equation*}
  \boldvtheta = \mathbf{o} + \mathbf{G}_1 \boldgamma_1 + \mathbf{G}_2 \boldgamma_2 + \mathbf{G}_3 \boldgamma_3 + \mathbf{G}_4 \boldgamma_4 + \tilde{\mathbf{u}}\,,
\end{equation*}
where $\boldvtheta = (\boldvtheta_1', \dots, \boldvtheta_\nt ')'$, $\mathbf{o} = (\mathbf{o}_1', \dots, \mathbf{o}_\nt ')'$, $\tilde{\mathbf{u}} = (\tilde{\mathbf{u}}_1', \dots, \tilde{\mathbf{u}}_\nt ')'$, $\boldgamma_1 = (\bolddelta_{01}^{* \prime}, \bolddelta_{02}^{* \prime}, \dots, \bolddelta_{0 \nt}^{* \prime})'$, $\boldgamma_2 = (\delta_{01}, \delta_{11}, \dots, \delta_{0 \nt}, \delta_{1 \nt})'$, $\boldgamma_3 = (\Tilde{\bolddelta}_0', \Tilde{\bolddelta}_1')'$, $\boldgamma_4 = (\overline{\delta}_0, \overline{\delta}_1, \boldalpha')'$. The observation matrices are $\mathbf{G}_1 = \II_{\ns \nt}$, 
\begin{equation*}
  \mathbf{G}_2 = \begin{bmatrix}
    \ones_\ns & \boldtilde{X}_{\, \bm{\cdot} 1} &  &  &\\
    & & \ones_\ns & \boldtilde{X}_{\, \bm{\cdot} 2} &  &  \\
    & &   &   \ddots & \ddots & \\
    & & & & \ones_\ns &  \boldtilde{X}_{\, \bm{\cdot} \nt}
  \end{bmatrix}\,, \qquad
  \mathbf{G}_3 = \begin{bmatrix}
    \II_\ns & \mbox{diag}(\boldtilde{X}_{\, \bm{\cdot} 1}) \\
     \vdots &  \vdots\\
    \II_\ns &  \mbox{diag}(\boldtilde{X}_{\, \bm{\cdot} \nt})
  \end{bmatrix}\,, \qquad
  \mathbf{G}_4 = \begin{bmatrix}
    \ones_\ns & \boldtilde{X}_{\, \bm{\cdot} 1} & {\mathbf{M}}_1\\
     \vdots &  \vdots & \vdots\\
    \ones_\ns &  \boldtilde{X}_{\, \bm{\cdot} \nt} & {\mathbf{M}}_\nt
  \end{bmatrix}\,.
\end{equation*}

Posterior samples are obtained using a Gibbs sampler with the following full conditionals, where the symbol ``$ \bullet $'' denotes data and all the parameters except for the parameter that is being updated.

{
\makeatletter
\@fleqntrue
\makeatother
\begin{enumerate}
  \item Sample regression coefficients, $\boldgamma_j$, for $j=1,2,3,4$:
  \begin{align*}
    & \boldgamma_j | \bullet \sim \Norm(\overline{\mathbf{V}}_j \overline{\boldmu}_j, \overline{\mathbf{V}}_j)\,,
  \end{align*}
  where $\overline{\mathbf{V}}_j = (\mathbf{K}_j + \mathbf{G}_j' \mathbf{G}_j / \tilde{\tau}_u^2)^{-1}$ and $\overline{\boldmu}_j = \mathbf{G}_j' \boldeta_j / \tilde{\tau}_u^2$, where $\boldeta_j = \boldvtheta - \mathbf{o} - \sum_{l \ne j}^{} \mathbf{G}_l \boldgamma_l$, $\mathbf{K}_j = \mathbf{H}_j' \mathbf{S}_j^{-1} \mathbf{H}_j$, and
  \begin{equation*}
    \mathbf{H}_1 = \begin{bmatrix}
      \mathbf{I}_\ns & & &\\
      -\mathbf{I}_\ns & \mathbf{I}_\ns & \\
       & \ddots & \ddots & \\
       & & -\mathbf{I}_\ns & \mathbf{I}_\ns \\
    \end{bmatrix}
    \quad \mbox{and} \quad
    \mathbf{S}_1^{-1} = \begin{bmatrix}
      \mathbf{V}_{\delta}^{* -1} & & & \\
      & \sigma_{w0}^{* -2}(\mathbf{D-W}) & &  \\
      & & \ddots &  \\
      & & & \sigma_{w0}^{* -2}(\mathbf{D-W}) \\
    \end{bmatrix}\,,
  \end{equation*}
  \begin{equation*}
    \mathbf{H}_2 = \begin{bmatrix}
      \mathbf{I}_2 & & & \\
      -\mathbf{I}_2 & \mathbf{I}_2 & \\
      & \ddots & \ddots & \\
      & & -\mathbf{I}_2 & \mathbf{I}_2
    \end{bmatrix}
    \quad \mbox{and} \quad
    \mathbf{S}_2^{-1} = \begin{bmatrix}
      \mbox{diag}(V_{\delta 0}^{-1}, V_{\delta 1}^{-1}) & & & \\
      & \mbox{diag}(\sigma_{w0}^{-2}, \sigma_{w1}^{-2}) & &  \\
      & & \ddots &  \\
      & & & \mbox{diag}(\sigma_{w0}^{-2}, \sigma_{w1}^{-2}) \\
    \end{bmatrix}\,,
  \end{equation*}
  \begin{equation*}
    \mathbf{H}_3 = \II_{2 \ns}
    \quad \mbox{and} \quad
    \mathbf{S}_3^{-1} = \begin{bmatrix}
      \sigma_{\delta 0}^{-2}(\mathbf{D-W}) & \\
      & \sigma_{\delta 1}^{-2}(\mathbf{D-W}) \\
    \end{bmatrix}\,,
  \end{equation*}
  \begin{equation*}
    \mathbf{H}_4 = \II_{\pmeas + 2}
    \quad \mbox{and} \quad
    \mathbf{S}_4^{-1} = 10^{-6} \II_{\pmeas + 2}\,.
  \end{equation*}

  \item Sample variance parameters of temporal effects, for $k=0,1$:
  \begin{align*}
    & \sigma_{wk}^2 | \bullet \sim \IG \left( a_{wk} + \frac{\nt-1}{2}, b_{wk} + \frac{1}{2} \sum_{t=2}^{\nt} (\delta_{kt} - \delta_{k, t-1})^2 \right)
  \end{align*}
  \item Sample variance parameters of spatial effects, for $k=0,1$:
  \begin{align*}
    & \sigma_{\delta k}^2 | \bullet \sim \IG \left( a_{\delta k} + \frac{\ns-1}{2}, b_{\delta k} + \frac{1}{2} \Tilde{\bolddelta}_k^\prime (\mathbf{D} -  \mathbf{W}) \Tilde{\bolddelta}_k \right)
  \end{align*}
  \item Sample variance parameter of spatio-temporal effect:
  \begin{align*}
    & \sigma_{w0}^{* 2} | \bullet \sim \IG \left( a_{w0}^* + \frac{(\ns-1)(\nt-1)}{2}, b_{w0}^* + \frac{1}{2} \sum_{t=2}^{\nt} (\bolddelta_{0t}^* - \bolddelta_{0, t-1}^*)^\prime (\mathbf{D} -  \mathbf{W}) (\bolddelta_{0t}^* - \bolddelta_{0, t-1}^*) \right)
  \end{align*}
  \item Sample the latent variables, $\vartheta_{it}$, for $i=1, \dots, \ns$ and $t=1, \dots, \nt$, using Metropolis-Hastings steps with a Gaussian proposal distribution:
  \begin{align*}
    & \vartheta_{it} | \bullet \propto \exp \left\{ Y_{it} \vartheta_{it} - \exp(\vartheta_{it}) - \frac{1}{2 \tilde{\tau}_u^2} (\vartheta_{it} - o_{it} - \vartheta_{it}^*)^2 \right\}\,,
  \end{align*}
  where $\vartheta_{it}^*$ is a generic element of $ \boldvtheta_{t}^* = (\vartheta_{11}^*, \vartheta_{21}^*, \dots, \vartheta_{\ns \nt}^*)' = \ones_\ns (\overline{\delta}_0 + \delta_{0t}) + \Tilde{\bolddelta}_0 + \bolddelta_{0t}^* + \boldtilde{X}_{t} (\overline{\delta}_1 + \delta_{1t}) + \boldtilde{X}_{t} \odot \Tilde{\bolddelta}_1 +  {\mathbf{M}}_t \boldalpha $.
  \item Sample the measurement error variance:
  \begin{align*}
    & \tilde{\tau}_u^2 | \bullet \sim \IG \left( a_u + \frac{\ns \nt}{2}, b_u + \frac{1}{2} \sum_{t=1}^{\nt} (\boldvtheta_t - \mathbf{o}_t - \boldvtheta_t^*)' (\boldvtheta_t - \mathbf{o}_t - \boldvtheta_t^*) \right)\,.
  \end{align*}
\end{enumerate}
}

\section{Details on Competing Models} \label{sec:tvc-details-competitors}
The proposed DGLM is compared to several approaches from the literature. These are fitted under the Bayesian paradigm, and the general hierarchical structure is as follows, for $i=1, \dots, \ns$ and $t=1, \dots, \nt$:
\begin{align*}
  Y_{it} | \vartheta_{it} &\sim \Poi(\exp(\vartheta_{it})) \\
  \vartheta_{it} &= o_{it} + \mathbf{A}_{it}' \mathbf{a}_i + \mathbf{C}_{it}' \mathbf{c}_i + \varepsilon_{it}, \quad \varepsilon_{it} \iid \Norm(0, \tau_\varepsilon^2) \\
  \mathbf{a}_i &\sim \Norm(\boldmu_a, \boldSigma_a) \\
  \boldmu_a &\sim \Norm(0, 10^6 \II_{p_a}) \\
  \boldSigma_a &\sim \IW(p_a + 1.1, 10^{-4} \II_{p_a}) \\
  \mathbf{c}_i &\sim \Norm(\zeros, 10^6 \II_{p_c}) \\
  \tau_\varepsilon^2 &\sim \IG(1, 2.1 \times 10^{-5})
\end{align*}
where $\mathbf{A}_{it}$ is a $p_a$-dimensional vector of covariates related to the exposure effect, $\mathbf{C}_{it}$ is a $p_c$-dimensional vector of covariates (i.e., intercept and measured confounders), and $\mathbf{a}_i$ and $\mathbf{c}_i$ are the corresponding vectors of coefficients. The vector $\boldmu_a$ and the matrix $\boldSigma_a$ are of main interest since they correspond to the overall exposure effect and the between-site variability, respectively.

The specification of $\mathbf{A}_{it}$ and $\mathbf{C}_{it}$ depends on the approach under consideration as listed below.
\begin{itemize}
  \item \textit{Null}: the model that does not account for unmeasured confounding, for which $\mathbf{A}_{it} = X_{it}$ and $\mathbf{C}_{it} = (1, {\MM}_{it}')'$. The regression coefficients are allowed to vary across areas but not in time.
  \item \textit{GLMadj}: the model that includes smooth function of time, $h(t; df)$, modelled as a natural cubic spline with $df$ degrees of freedom. Let $\mathbf{B}_t$ denote the vector of basis functions evaluated at time $t$. Then, we define $\mathbf{A}_{it} = X_{it}$ and $\mathbf{C}_{it} = (1, {\MM}_{it}', \mathbf{B}_t')'$.
  \item \textit{Dummy}: \citet{dai2014associations,klompmaker2023effects} assume that the long-term exposure--outcome association vary by season. Following also \citet{chen2013seasonal,peng2005seasonal}, we may extend this approach to our setting by including interaction terms. Let $D_{jt}$ indicate an indicator variable for the $j$-th calendar season, for $j=1, \dots, 4$, that is:
  \begin{equation*}
    D_{jt} = \begin{cases}
      1 & \mbox{if the day $t$ is in the $j$-th season} \\
      0 & \mbox{otherwise}
    \end{cases}
  \end{equation*}
  Then, $\mathbf{A}_{it} = (X_{it} D_{1t}, \dots, X_{it} D_{4t})'$ and $\mathbf{C}_{it} = (1, { \MM}_{it}', \mathbf{B}_t')'$. Henceforth, the Dummy model is similar to GLMadj, but it also captures season-specific exposure effects.
  \item \textit{Periodic}: the exposure effect is assumed to be a linear combination of trigonometric functions, as in \citet{peng2005seasonal,chen2013seasonal}. In this case, $\mathbf{A}_{it} = (X_{it}, X_{it} \cos(2 \pi t / 365), X_{it} \sin(2 \pi t / 365))'$ and $\mathbf{C}_{it} = (1, {\MM}_{it}', \mathbf{B}_t')'$.
  \item \textit{JDZ}: the model proposed by \citet{janes2007trends}, where the local exposure effect is of interest, hence $\mathbf{A}_{it} = (X_{it} - \widehat{X}_{it})'$ and $\mathbf{C}_{it} = (1, \widehat{X}_{it}, {\MM}_{it}', \mathbf{B}_t^{* \prime})'$. Here, $\widehat{X}_{it}$ denotes the overall temporal trend (see also Section 3), and $\mathbf{B}_t^*$ denotes the vector of basis functions from a natural cubic spline with $df-1$ degrees of freedom, evaluated at time $t$.
\end{itemize}

The unknown function $h(\cdot;\cdot)$ is modelled as a natural cubic spline of time with $df=43$ degrees of freedom (i.e., 8 equally-spaced knots per year plus 3 more knots to account for the peaks in the SMR time series), and is used to control for unmeasured confounding. The overall trend, $\widehat{X}_t$, is obtained as the fitted values of a regression of $X_{it}$ on a natural cubic spline of time with $4$ degrees of freedom per year. In the real data application, bases from natural cubic splines of temperature (with 7 degrees of freedom) and of relative humidity (with 4 degrees of freedom) are included as measured confounders. ${\MM}_{it}$ also includes the indicator variables for the day of the week, for holidays, and for the lockdown periods.

\section{Additional Details on Simulation Study 1} \label{sec:tvc-simulations-additional}

\subsection{Data Generation Mechanism} \label{subsec:tvc-dgm}

To simulate the variables, we use the following data generation mechanism. The spatial domain comprises $\ns=117$ contiguous HDs, while the temporal dimension is set to $\nt=600$. Let $\XX = (\XX_1, \dots, \XX_{\nt})'$, with $\XX_t = (X_{1t}, \dots, X_{\ns t})'$, $t=1,\dots,\nt$, be the exposure, and let $\ZZ = (\ZZ_1, \dots, \ZZ_{\nt})'$, with $\ZZ_t = (Z_{1t}, \dots, Z_{\ns t})'$, $t=1,\dots,\nt$, be the unmeasured confounder. It is thus assumed that
\begin{equation} \label{eq:joint-xz}
  \begin{bmatrix}
      \mathbf{X} \\
      \mathbf{Z}
  \end{bmatrix} \sim \Norm \left(\begin{bmatrix}
      \boldmu_x \\
      \boldmu_z
  \end{bmatrix} , \begin{bmatrix}
      \boldSigma_{x} & \boldSigma_{xz} \\
      \boldSigma_{zx} & \boldSigma_{z}
  \end{bmatrix} \right)\,,
\end{equation}
where $\boldmu_x$ and $\boldmu_z$ are the spatio-temporal means generated from Fourier basis expansions. It is assumed that the covariance matrices can be factorized as $\boldSigma_{x} = \boldSigma_x ^\half \boldSigma_x ^ \halft$, $\boldSigma_{z} = \boldSigma_z ^\half \boldSigma_z ^ \halft$, $\boldSigma_{xz} = \boldSigma_{zx}' = \rho_{xz} \boldSigma_x ^\half \boldSigma_z ^ \halft$, and $\rho_{xz} = \Cor (\mathbf{X}, \mathbf{Z})$.

We then assume that exposure and unmeasured confounder are spatial VAR(1) processes \citep{cressie2015statistics,sahu2022bayesian}. We now explain how $\boldSigma_{x}$ is parameterized, and the same applies to $\boldSigma_{z}$. If a stable VAR(1) model is assumed for the exposure where the temporal autocorrelation is controlled by a scalar, $\phi_x^T \in (0,1)$, then it can be shown \citep[see, for example,][Section 2.1.4]{lutkepohl2005new} that $\boldSigma_x = \boldGamma_{x,ar} \otimes {\bf C}_x(0)$, where 
\begin{equation*}
  {\bf C}_x(0) = \E \left[ \XX_t \XX_t' \right] = \frac{\tau_x^2}{1 - (\phi_x^T)^2} \boldOmega_{x,pcar}^{-1}
\end{equation*}
indicates a zero-lag covariance matrix, and $\boldGamma_{x,ar}$ is the correlation matrix of a first-order autoregressive process with generic element $ \left( \boldGamma_{x,ar} \right)_{jk} = \left( {\phi_x^T} \right)^{|j-k|} $. The spatial dependence is defined by the matrix $\boldOmega_{x,pcar}$, which has a proper conditional autoregressive (PCAR) specification \citep{banerjee2014hierarchical,besag1974spatial,sahu2022bayesian}:
\begin{equation} \label{eq:pcar-model}
  \boldOmega_{x,pcar} = \DD - \phi_x^S \WW\,,
\end{equation}
where $\DD$ is a diagonal matrix with the number of neighbors of each HD on the diagonal, $\WW$ is the binary adjacency matrix, and $\phi_x^S \in (0,1)$ is the spatial autocorrelation parameter. $\WW$ is an $\ns \times \ns$ matrix with entries $w_{ij}=1$ if areal units $i$ and $j$ share a border, and $w_{ij}=0$ otherwise; also, $w_{ii}=0$.

We set $\rho_{xz}=0.5$, 
$\tau_x^2 = 4 \times 10^{-4}$, and $\tau_z^2 = 1 \times 10^{-3}$, and then use different values for the spatial and temporal autocorrelation parameters to define the scenarios in Table 1 of the main manuscript.

Once all the matrices in Equation (\ref{eq:joint-xz}) are defined, we simulate the exposure from its marginal distribution, then the unmeasured confounder from the conditional distribution $\ZZ | \XX \sim \Norm \left( \boldmu_{z|x}, \boldSigma_{z|x} \right)$ where:
\begin{align*}
  \boldmu_{z|x} &= \E \left[\ZZ|\mathbf{X=x}\right] = \boldmu_z + \boldSigma_{zx} \boldSigma_{x}^{-1} (\mathbf{x} - \boldmu_x), \\
  \boldSigma_{z|x} &= \Var \left[\mathbf{Z}|\mathbf{X=x}\right] = \boldSigma_{z} - \boldSigma_{zx} \boldSigma_{x}^{-1} \boldSigma_{xz}\,.
\end{align*} 

\noindent Finally, the outcome is generated using the following model:
\begin{align}
  Y_{it}|\mu_{it} &\stackrel{ind}{\sim} \Poi (\mu_{it}) \label{eq:tvc-generate-outcome} \\
  \vartheta_{it} &= o_{it} + \beta_{0} + \beta_{1it} X_{it} + Z_{it} + u_{it}, \quad u_{it} \iid \Norm(0, \tau_u^2)\,, \label{eq:tvc-generate-linpred1}
\end{align}
where $\mu_{it} = \exp(\vartheta_{it})$, $o_{it} = \log (E_{it})$ is the offset term, and $\tau_u^2= 10^{-3}$. The expected counts, $E_{it}$, are obtained by using the resident population aged $\ge 65$ years and the age-specific standardized mortality rates of year 2018. The population and the rates are assumed to be constant in time.

\clearpage

\subsection{Performance Assessment} \label{subsec:tvc-mse}

To complete the results of Simulation Study 1, Tables \ref{tab:tvc-MSE-3scenarios}--\ref{tab:tvc-MAE-3scenarios} report, for each scenario and model, the value for the three indices, namely:
\begin{itemize}
  \item average mean squared error, $MSE_{avg} = \frac{1}{\ns \nt} \sum_{i=1}^{\ns} \sum_{t=1}^{\nt}  \left[ \frac{1}{100 } \sum_{j=1}^{100} \left( \E^{(j)}[P_{it}|\bullet] - P_{it} \right)^2 \right] $, where $\E^{(j)}[P_{it}|\bullet]$ is the posterior mean for the $j$-th simulated dataset;
  \item maximum absolute bias, $MAB =  \max_{i,t} \frac{1}{100 } \sum_{j=1}^{100} | \E^{(j)}[P_{it}|\bullet] - P_{it}| $
  \item average mean absolute error, $MAE_{avg} = \frac{1}{\ns \nt} \sum_{i=1}^{\ns} \sum_{t=1}^{\nt} \left[ \frac{1}{100} \sum_{i=1}^{100} |\E^{(j)}[P_{it}|\bullet] - P_{it}| \right]$
\end{itemize}
Here, the differences are computed with respect to $P_{it} = 100(\exp \{10 (\beta_{1it}) \}-1) $, i.e. the true simulated percent change of risk, which varies spatio-temporally.

Tables \ref{tab:tvc-MSE-3scenarios}--\ref{tab:tvc-MAE-3scenarios} indicate that, SDGLMC is the best model among those compared in terms of $MSE_{avg}$, $MAB$, and $MAE_{avg}$ for all scenarios. Recalling the decomposition of $\beta_{1it}$ in Section 3.1 of the main manuscript, we have also computed the average mean squared error for each of the components of the exposure effect -- baseline, temporal and spatial. SDGLMC provides the lowest values in most of the cases, and the improvements over the other approaches are more pronounced for the temporal component.

Since we want the overall error to be as low as possible, and since the shape of the exposure effect is not known in practice, simulation results suggest that the proposed model is the most robust choice across all scenarios.

\begin{table}[!htbp]
  \renewcommand{\arraystretch}{1.2}
  \caption[Simulation Study 1: $MSE_{avg}$ for each Scenario]{Simulation Study 1: $MSE_{avg}$ calculated for each model considered under each Scenario. The lowest values are indicated in italic.}
  \label{tab:tvc-MSE-3scenarios}
  \centering
  \begin{tabular}{lllcccccc}
    \hline
    Scenario & Component & Formula & SDGLMC & Null & GLMadj   & Dummy & Periodic & JDZ  \\
    \hline
    \multirow{4}{*}{S1} %
    & Overall & $100(\exp \{10 \beta_{1it} \}-1) $ & \textit{0.669} & 36.214 & 2.417   & 1.888 & 2.780 & 2.522  \\
    & Baseline & $100(\exp \{10 \overline{\delta}_1 \}-1) $ & \textit{0.115} & 34.568 & 0.193    & 0.142 & 0.225 & 0.275 \\
    & Temporal & $100(\exp \{10 \delta_{1t} \}-1) $ & \textit{0.200} & 1.418 & 1.418   & 0.889 & 1.705 & 1.418  \\
    & Spatial & $100(\exp \{10 \tilde{\delta}_{1i} \}-1) $ & 0.354 & \textit{0.228} & 0.806    & 0.855 & 0.844 & 0.829  \\
    \hline
    \multirow{4}{*}{S2} %
    & Overall & $100(\exp \{10 \beta_{1it} \}-1) $  & \textit{5.617} & 20.440 & 6.596   & 6.022 & 8.118 & 6.681 \\
    & Baseline & $100(\exp \{10 \overline{\delta}_1 \}-1) $  & 5.246 & 13.331 & \textit{4.807}    & 4.897 & 6.568 & 4.893 \\
    & Temporal & $100(\exp \{10 \delta_{1t} \}-1) $  & \textit{0.103} & 1.418 & 1.418   & 0.837 & 1.204 & 1.418 \\
    & Spatial & $100(\exp \{10 \tilde{\delta}_{1i} \}-1) $  & \textit{0.268} & 5.691 & 0.371    & 0.277 & 0.288 & 0.369 \\
    \hline
    \multirow{4}{*}{S3} %
    & Overall & $100(\exp \{10 \beta_{1it} \}-1) $ & \textit{0.857} & 31.247 & 2.286   & 1.699 & 2.222 & 2.337 \\
    & Baseline & $100(\exp \{10 \overline{\delta}_1 \}-1) $ & 0.509 & 29.444 & 0.370    & \textit{0.310} & 0.423 & 0.417 \\
    & Temporal & $100(\exp \{10 \delta_{1t} \}-1) $ & \textit{0.133} & 1.418 & 1.418   & 0.886 & 1.281 & 1.418 \\
    & Spatial & $100(\exp \{10 \tilde{\delta}_{1i} \}-1) $ & \textit{0.216} & 0.385 & 0.498    & 0.500 & 0.506 & 0.502 \\
    \hline
  \end{tabular}
\end{table}

\begin{table}[!htbp]
  \renewcommand{\arraystretch}{1.2}
  \caption[Simulation Study 1:  $MAB$ for each Scenario]{Simulation Study 1:  $MAB$ calculated for each model considered under each Scenario. The lowest values are indicated in italic.}
  \label{tab:tvc-MAB-3scenarios}
  \centering
  \begin{tabular}{lllcccccc}
    \hline
    Scenario & Component & Formula & SDGLMC & Null & GLMadj   & Dummy & Periodic & JDZ  \\
    \hline
    \multirow{4}{*}{S1} %
    & Overall & $100(\exp \{10 \beta_{1it} \}-1) $ &  \textit{1.212} & 7.770 & 3.816    & 3.796 & 4.010 & 3.618 \\
    & Baseline & $100(\exp \{10 \overline{\delta}_1 \}-1) $ &  0.295 & 5.856 & 0.189   & \textit{0.022} & 0.340 & 0.316 \\
    & Temporal & $100(\exp \{10 \delta_{1t} \}-1) $ & \textit{0.276} & 2.100 & 2.100    & 1.913 & 2.497 & 2.100 \\
    & Spatial & $100(\exp \{10 \tilde{\delta}_{1i} \}-1) $ &  \textit{0.934} & 1.373 & 1.859    & 1.873 & 1.787 & 1.788 \\
    \hline
    \multirow{4}{*}{S2} %
    & Overall & $100(\exp \{10 \beta_{1it} \}-1) $  & \textit{2.898} & 5.896 & 3.958    & 4.048 & 5.117 & 3.904 \\
    & Baseline & $100(\exp \{10 \overline{\delta}_1 \}-1) $  & 2.280 & 3.647 & \textit{2.183}    & 2.204 & 2.555 & 2.203 \\
    & Temporal & $100(\exp \{10 \delta_{1t} \}-1) $  &  \textit{0.573} & 2.100 & 2.100    & 2.003 & 2.190 & 2.100 \\
    & Spatial & $100(\exp \{10 \tilde{\delta}_{1i} \}-1) $  & 1.198 & 1.095 & 0.387    & 0.356 & 0.372 & \textit{0.313} \\
    \hline
    \multirow{4}{*}{S3} %
    & Overall & $100(\exp \{10 \beta_{1it} \}-1) $ & \textit{1.506} & 7.683 & 2.767    & 2.769 & 3.607 & 2.791 \\
    & Baseline & $100(\exp \{10 \overline{\delta}_1 \}-1) $ & 0.599 & 5.413 & 0.479   & \textit{0.403} & 0.529 & 0.523 \\
    & Temporal & $100(\exp \{10 \delta_{1t} \}-1) $ & \textit{0.211} & 2.100 & 2.100    & 1.968 & 2.250 & 2.100 \\
    & Spatial & $100(\exp \{10 \tilde{\delta}_{1i} \}-1) $ & \textit{0.740} & 1.691 & 1.100    & 1.024 & 1.114 & 1.168 \\
    \hline
  \end{tabular}
\end{table}

\begin{table}[!htbp]
  \renewcommand{\arraystretch}{1.2}
  \caption[Simulation Study 1: $MAE_{avg}$ for each Scenario]{Simulation Study 1: $MAE_{avg}$ calculated for each model considered under each Scenario. The lowest values are indicated in italic.}
  \label{tab:tvc-MAE-3scenarios}
  \centering
  \begin{tabular}{lllcccccc}
    \hline
    Scenario & Component & Formula & SDGLMC & Null & GLMadj   & Dummy & Periodic & JDZ  \\
    \hline
    \multirow{4}{*}{S1} %
    & Overall & $100(\exp \{10 \beta_{1it} \}-1) $ &  \textit{0.654} & 5.856 & 1.287 & 1.111 & 1.330 & 1.315 \\
    & Baseline & $100(\exp \{10 \overline{\delta}_1 \}-1) $ & 0.297 & 5.856 & 0.340 & \textit{0.286} & 0.400 & 0.422\\
    & Temporal & $100(\exp \{10 \delta_{1t} \}-1) $ &  \textit{0.355} & 1.065 & 1.065 & 0.801 & 1.046 & 1.065 \\
    & Spatial & $100(\exp \{10 \tilde{\delta}_{1i} \}-1) $ & 0.468 & \textit{0.374} & 0.728 & 0.751 & 0.744 & 0.739 \\
    \hline
    \multirow{4}{*}{S2} %
    & Overall & $100(\exp \{10 \beta_{1it} \}-1) $  &  2.281 & 3.868 & 2.226 & \textit{2.213} & 2.564 & 2.242 \\
    & Baseline & $100(\exp \{10 \overline{\delta}_1 \}-1) $  & 2.280 & 3.647 & \textit{2.183} & 2.204 & 2.555 & 2.203 \\
    & Temporal & $100(\exp \{10 \delta_{1t} \}-1) $  & \textit{0.253} & 1.065 & 1.065 & 0.794 & 0.823 & 1.065 \\
    & Spatial & $100(\exp \{10 \tilde{\delta}_{1i} \}-1) $  & \textit{0.402} & 1.908 & 0.485 & 0.415 & 0.425 & 0.485 \\
    \hline
    \multirow{4}{*}{S3} %
    & Overall & $100(\exp \{10 \beta_{1it} \}-1) $ & \textit{0.735} & 5.413 & 1.272 & 1.060 & 1.176 & 1.285 \\
    & Baseline & $100(\exp \{10 \overline{\delta}_1 \}-1) $ & 0.599 & 5.413 & 0.479 & \textit{0.404} & 0.529 & 0.523 \\
    & Temporal & $100(\exp \{10 \delta_{1t} \}-1) $ &  \textit{0.287} & 1.065 & 1.065 & 0.806 & 0.907 & 1.065 \\
    & Spatial & $100(\exp \{10 \tilde{\delta}_{1i} \}-1) $ & \textit{0.365} & 0.495 & 0.557 & 0.556 & 0.561 & 0.558 \\
    \hline
  \end{tabular}
\end{table}

\clearpage

\section{Simulation Study 2}
We now illustrate \textit{Simulation Study 2}, where we want to assess whether SDGLMC is able to recover the first derivative of the unknown functional form, which represents the marginal effect of the exposure on the outcome. To this end, we consider the Scenario S1 again, but we assume that the linear predictor is as follows:
\begin{equation*}
  \vartheta_{it} = o_{it} + \beta_{0} + f(X_{it}; \boldgamma_f) + Z_{it} + u_{it}\,,
\end{equation*}
where $f(X_{it}; \boldgamma_f) = \gamma_{f0} + \gamma_{f1} X_{it} + \gamma_{f2} X_{it}^2 + \gamma_{f3} X_{it}^3 $, $\gamma_{f0}=-0.08$, $\gamma_{f1}=0.02$, $\gamma_{f2}=-8 \times 10^{-5}$, $\gamma_{f3}=5 \times 10^{-8}$. Henceforth, the true values of the time-varying coefficients are $\beta_{1t} = 0.02 - 1.6 \times 10^{-4} \widehat{X}_{it} + 1.5 \times 10^{-7} \widehat{X}_{it}^2$, and $\beta_{0it} = \beta_{0} -0.08 + 0.02 \widehat{X}_{it} - 8 \times 10^{-5} \widehat{X}_{it}^2 + 5 \times 10^{-8} \widehat{X}_{it}^3 + Z_{it}$. All the other parameters are the same as in the previous Section.

Table \ref{tab:tvc-indices-dglms} reports the $MSE_{avg}$, $MAB$, and $MAE_{avg}$ that were obtained. The small values of the indices indicate that the proposed model is able to recover the first derivative of the unknown functional form. Some confounding bias is still present, but it is not attributable to the Taylor's series approximation, since it is not present when the data are generated without unmeasured confounder (i.e., $Z_{it}=0$).

\begin{table}[!htbp]
  \renewcommand{\arraystretch}{1.2}
  \caption[Simulation Study 2: $MSE_{avg}$, $MAB$, and $MAE_{avg}$]{Simulation Study 2: $MSE_{avg}$, $MAB$, and $MAE_{avg}$ calculated for SDGLMC and SDGLMC\_single under Scenario S1 with $f(X_{it}; \boldgamma_f) = \gamma_{f0} + \gamma_{f1} X_{it} + \gamma_{f2} X_{it}^2 + \gamma_{f3} X_{it}^3 $. The lowest values for each index are indicated in italic.}
  \label{tab:tvc-indices-dglms}
  \centering
  \begin{tabular}{lcccccc}
    \hline
    \multirow{2}{*}{~} & \multicolumn{2}{c}{$MSE_{avg}$} & \multicolumn{2}{c}{$MAB$} & \multicolumn{2}{c}{$MAE_{avg}$} \\
    \cline{2-7}
    Component & SDGLMC & SDGLMC\_single & SDGLMC & SDGLMC\_single & SDGLMC & SDGLMC\_single \\
    \hline
    Overall & \textit{1.381} & 1.897 & \textit{2.288} & 3.330 & \textit{0.929} & 1.094 \\
    Baseline & \textit{0.050} & 0.398 & \textit{0.170} & 0.614 & \textit{0.185} & 0.614 \\
    Temporal & \textit{0.376} & 0.503 & \textit{2.000} & 2.579 & \textit{0.481} & 0.544 \\
    Spatial & \textit{0.945} & 0.984 & \textit{0.452} & 0.860 & \textit{0.762} & 0.741 \\
    \hline
  \end{tabular}
\end{table}

\subsection{On The Removal of Large-Scale Spatio-Temporal Trend.} \label{subsec:tvc-removal-trend}
We argue that the term $X_{it} - \widehat{X}_{it}$, for $i=1, \dots, \ns$ and $t=1, \dots, \nt$, should be used in place of simply $X_{it}$ in the DGLM. Removing the large-scale spatio-temporal trend from the exposure not only follows from theoretical arguments (see Equation (3) in the main paper), but also it is crucial to control for confounding bias. The local variations in the exposure are in fact less likely to be confounded by the unmeasured confounder, which is assumed to be spatially and temporally smooth \citep{janes2007trends,greven2011approach} -- see also Section 3 of the manuscript. To illustrate this with simulated data, we compare the proposed SDGLMC with a reparameterized model that does not remove the large-scale trend from the exposure. Essentially, if the first-order Taylor's series is evaluated at $0$, Equation (5) in the paper can be rewritten as:
\begin{equation*}
  \vartheta_{it} = o_{it} + \beta_{0it}^* + \beta_{1t}^* X_{it} + {\mathbf{M}}_{it}' \boldalpha + \tilde{u}_{it}\,,
\end{equation*}
where $\beta_{1it}^*=f'(0; \boldgamma_f)$ and $\beta_{0it}^* = \beta_{0} + f(0; \boldgamma_f) + Z_{it}$.

We refer to this model as \textit{SDGLMC\_single}, since it is a single-step approach that does not require to estimate the large-scale spatio-temporal trend.

The results for the reparameterized approach are reported in Table \ref{tab:tvc-indices-dglms}, where it is confirmed that SDGLMC outperforms SDGLMC\_single according to all indices.

\section{Application: Further Analyses}

\subsection{Comparison with Restricted Model} \label{sec:tvc-restricted-model}
\begin{figure}[!ht]
  \centering
  \includegraphics[width=15cm]{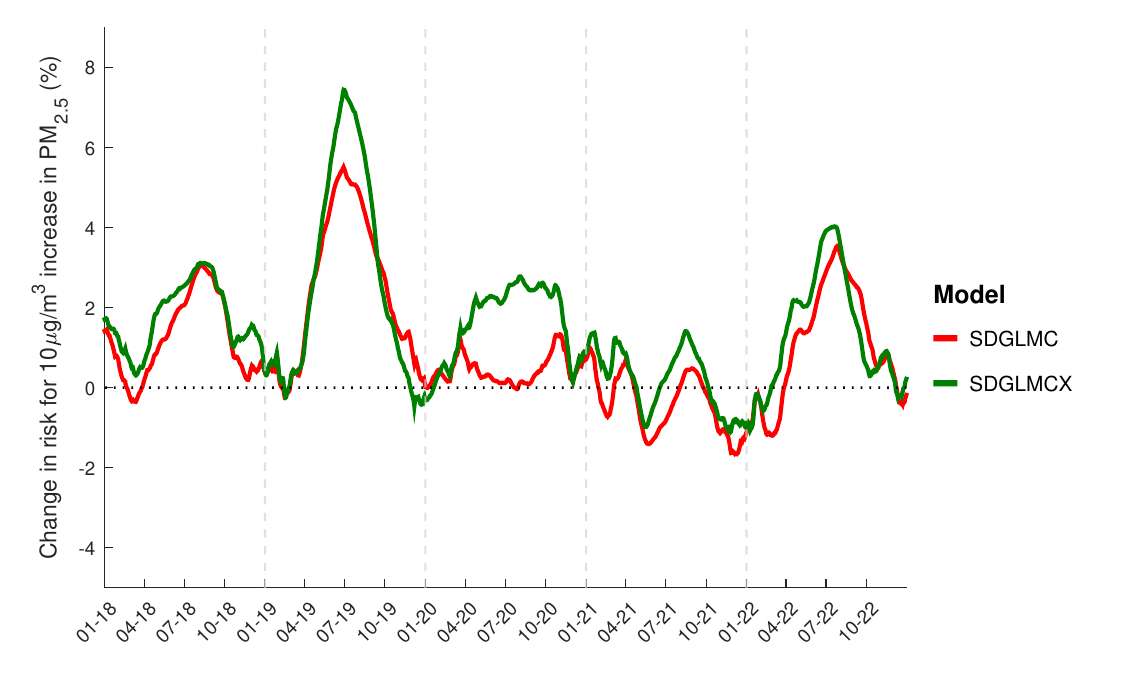}
  \caption[DGLM and SDGLMCX]{Estimated time-varying percent change of risk associated with a $10\ \mu g/m^3$ increase in the \pmtwopfive\ concentrations. The red line represents the posterior mean of SDGLMC, while the green line the posterior mean of SDGLMCX.}
  \label{fig:tvc-application-beta1t-DGLM-and-DGLM-X}
\end{figure}

We assess the impact that imposing some constraints on the proposed approach has on the estimated exposure effect. Specifically, we remove the measured confounders from the model specification. This is equivalent to imposing that $\boldalpha=\zeros$ in Equation (6) of the main text. Hence, the linear predictor is specified as follows:
\begin{equation*}
	\boldvtheta_{t} = \mathbf{o}_t + \ones_\ns (\overline{\delta}_0 + \delta_{0t}) + \Tilde{\bolddelta}_0 + \bolddelta_{0t}^* + %
  \boldtilde{X}_{t} (\overline{\delta}_1 + \delta_{1t}) + \boldtilde{X}_{t} \odot \Tilde{\bolddelta}_1 +%
  \tilde{\uu}_{t}\,.
\end{equation*}
We refer to the implied model as SDGLMCX to emphasize that the exposure is the only covariate. Figure \ref{fig:tvc-application-beta1t-DGLM-and-DGLM-X} shows the posterior mean of the percent change of risk, as defined above, under the SDGLMCX (in green) and the proposed DGLM (in red).
It suggests that, in general, the temporal dynamic of the exposure effect would not be affected much if the measured confounders were not included in the model. Henceforth, the proposed approach can provide accurate estimates even in complex scenarios. 

However, it is worth noting that SDGLMCX provides greater estimates than SDGLMC during the summer, while during the winter the estimates are quite similar. This suggests that the confounding effect may change over time. We hypothesize that this is due to air temperature exceeding a certain threshold; in other words, confounding may occur due to heat waves.
In fact, heat waves are demonstrated to have an increased effect on mortality during summer days with relatively high levels of \pmtwopfive\ concentrations, and this positive correlation may be the source of the overestimation \citep{analitis2008investigating,analitis2014effects}. Another study observed a similar result for the population aged 75 years or more \citep{analitis2018synergistic}. Another possible explanation for the overestimation could be the modification of the \pmtwopfive\ chemical composition, with high levels of the most toxic compounds developing in summer \citep{vinson2020seasonal}.

\subsection{Comparison with a GLM with interactions} \label{sec:tvc-application-GLMint}
We argue that the summer peaks of the exposure effect are likely due to potential interactions between \pmtwopfive\ and temperature or other confounders. To provide further evidence, we compare the proposed approach with a model that includes an interaction term between \pmtwopfive and each measured confounder. Therefore, we consider a hierarchical model, referred to as \textit{GLMint}, that follows the same structure as defined in Section \ref{sec:tvc-details-competitors}, but with $\mathbf{A}_{it} = (X_{it}, X_{it}{\MM}_{it}')'$ and $\mathbf{C}_{it} = (1, {\MM}_{it}', \mathbf{B}_t')'$.

Given the hierarchical structure of GLMint, the exposure effect can be estimated for each HD by summing the \pmtwopfive\ coefficient and the interaction coefficients (multiplied by the confounder values measured in the respective HD at each point in time). The estimates are then averaged across the HDs to obtain a single estimate for each time point.

\begin{figure}[!ht]
  \centering
  \includegraphics[width=15cm]{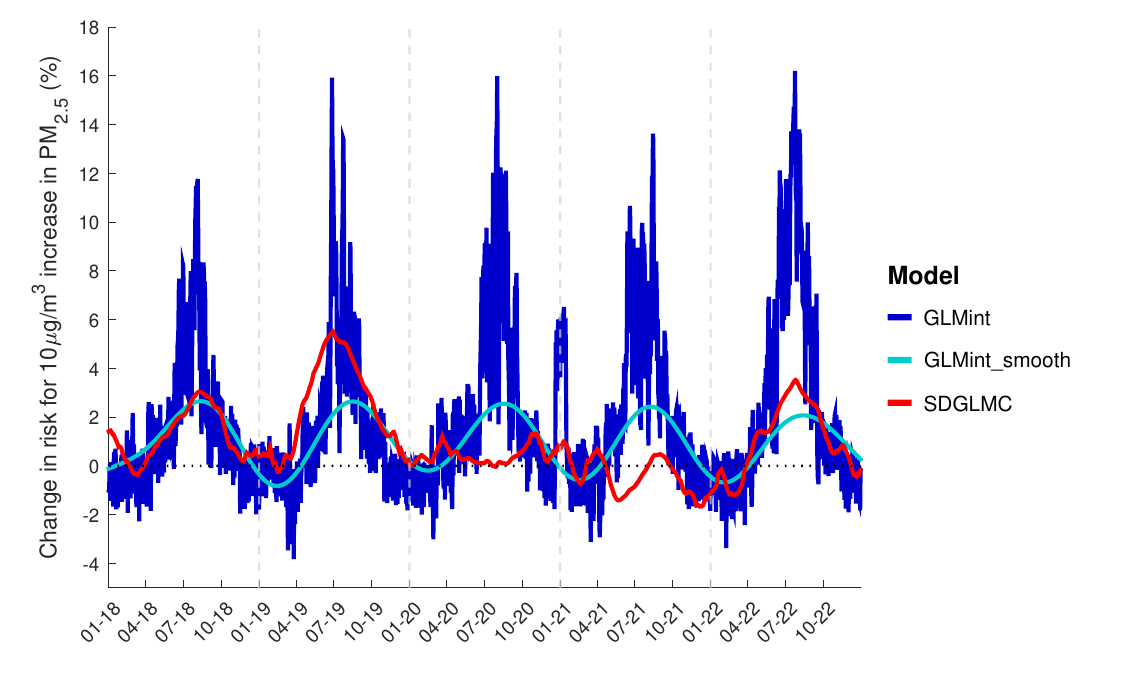}
  \caption[GLMint and DGLM]{Estimated time-varying percent change of risk associated with a $10\ \mu g/m^3$ increase in the \pmtwopfive\ concentrations.}
  \label{fig:tvc-application-beta1t-GLMint}
\end{figure}

Figure \ref{fig:tvc-application-beta1t-GLMint} shows the posterior mean of the percent change of risk (as defined in previous sections) under SDGLMC (in red) and GLMint (in dark blue). The estimates from both models appear similar during winter. However, GLMint may overestimate the exposure effect during summer, with peaks approximately equal to $16\%$, which is unexpected for short-term associations. This suggests that GLMint is not able to capture the temporal dynamics of the exposure effect as well as SDGLMC. 

Given that the effect estimated by GLMint is not smooth in time, another hierarchical model, \textit{GLMint\_smooth}, is added to the comparison. It includes interaction terms between \pmtwopfive\ and smoothed versions of temperature and relative humidity, and is depicted in light blue in Figure \ref{fig:tvc-application-beta1t-GLMint}. The smoothed confounders are obtained by using a principal spline with $15$ spatial bases and $15$ temporal bases. The results show that the temporal dynamics under GLMint\_smooth and SDGLMC are similar in 2018 and 2022. However, it seems that GLMint\_smooth could underestimate the effect in 2019, and could overestimate it during the pandemic. Furthemore, GLMint\_smooth and the Periodic model seem to provide similar estimates (see Figure 3 in the main manuscript).

Finally, this analysis corroborates the hypothesis of effect modification by unmeasured spatio-temporal confounders, which is not well captured by GLMint or GLMint\_smooth. The results indicate that the possibility of a more complex interaction between \pmtwopfive\ and the confounders cannot be ruled out.

\end{document}